\documentclass[iop, apj, twocolappendix, numberedappendix]{emulateapj}

\newif\iflocal
\localfalse

\usepackage{graphics}
\usepackage{graphicx}
\usepackage{amssymb,amsmath}
\usepackage{color}
\usepackage[colorlinks=true,urlcolor=black,linkcolor=blue,citecolor=blue]{hyperref}
\usepackage{xspace}
\usepackage{sistyle}
\SIthousandsep{,}
\usepackage{natbib}

\iflocal

\newcommand{\kpch}{\>{h^{-1}{\rm kpc}}}
\newcommand{\mpch}{\>h^{-1}{\rm {Mpc}}}
\newcommand{\gpch}{\>h^{-1}{\rm {Gpc}}}

\newcommand{\msunh}{\>h^{-1} M_\odot}

\newcommand{\cm}{\>{\rm cm}}

\def\gcm3{\mathrm{g} / \mathrm{cm}^3}

\def\LCDM{$\Lambda$CDM\xspace}

\def\mvir{M_{\rm vir}}
\def\rvir{R_{\rm vir}}
\def\cvir{c_{\rm vir}}

\def\rtom{R_{\rm 200m}}
\def\ctom{c_{\rm 200m}}

\def\rtoc{R_{\rm 200c}}
\def\ctoc{c_{\rm 200c}}

\def\nutoc{\nu_{\rm 200c}}

\def\cfoc{c_{\rm 500c}}

\def\mdelta{M_{\Delta}}
\def\rdelta{R_{\Delta}}
\def\cdelta{c_{\Delta}}

\def\rhoc{\rho_{\rm c}}
\def\rhom{\rho_{\rm m}}
\def\rhoref{\rho_{\rm ref}}

\def\rs{r_{\rm s}}

\def\colossus{\textsc{Colossus}\xspace}

\def\gtsima{$\; \buildrel > \over \sim \;$}
\def\ltsima{$\; \buildrel < \over \sim \;$}
\def\prosima{$\; \buildrel \propto \over \sim \;$}
\def\gsim{\lower.7ex\hbox{\gtsima}}
\def\lsim{\lower.7ex\hbox{\ltsima}}
\def\simgt{\lower.7ex\hbox{\gtsima}}
\def\simlt{\lower.7ex\hbox{\ltsima}}
\def\simpr{\lower.7ex\hbox{\prosima}}

\else

\fi
\def\cm{$c$--$M$\xspace}
\def\cnu{$c$--$\nu$\xspace}
\def\cmr{$c$--$M$ relation\xspace}
\def\cnur{$c$--$\nu$ relation\xspace}
\def\cmrs{$c$--$M$ relations\xspace}

\def\rdelta{R_{\Delta}}
\def\mdelta{M_{\Delta}}
\def\rs{r_{\rm s}}

\def\prho{P_{\rho}}
\def\pmass{P_{\rm m}}

\def\rhoc{\rho_{\rm c}}
\def\rhom{\rho_{\rm m}}
\def\rhos{\rho_{\rm s}}
\def\rhoref{\rho_{\rm ref}}

\def\mpe{M_{\rm pe}}
\def\zpe{z_{\rm pe}}
\def\cpe{c_{\rm pe}}
\def\nupe{\nu_{\rm pe}}
\def\nucut{\nu_{\rm cut}}

\def\neff{n_{\rm eff}}
\def\alphaeff{\alpha_{\rm eff}}

\def\cvir{c_{\rm vir}}

\def\rtom{R_{\rm 200m}}
\def\ctom{c_{\rm 200m}}

\def\rtoc{R_{\rm 200c}}
\def\ctoc{c_{\rm 200c}}
\def\nutoc{\nu_{\rm 200c}}

\def\cfoc{c_{\rm 500c}}

\usepackage{etoolbox}
\makeatletter
\patchcmd{\NAT@citex}
  {\@citea\NAT@hyper@{\NAT@nmfmt{\NAT@nm}\NAT@date}}
  {\@citea\NAT@nmfmt{\NAT@nm}\NAT@hyper@{\NAT@date}}
  {}
  {}
\patchcmd{\NAT@citex}
  {\@citea\NAT@hyper@{%
     \NAT@nmfmt{\NAT@nm}%
     \hyper@natlinkbreak{\NAT@aysep\NAT@spacechar}{\@citeb\@extra@b@citeb}%
     \NAT@date}}
  {\@citea\NAT@nmfmt{\NAT@nm}%
   \NAT@aysep\NAT@spacechar%
   \NAT@hyper@{\NAT@date}}
  {}
  {}
\patchcmd{\NAT@citex}
  {\@citea\NAT@hyper@{%
     \NAT@nmfmt{\NAT@nm}%
     \hyper@natlinkbreak{\NAT@spacechar\NAT@@open\if*#1*\else#1\NAT@spacechar\fi}%
       {\@citeb\@extra@b@citeb}%
     \NAT@date}}
  {\@citea\NAT@nmfmt{\NAT@nm}%
   \NAT@spacechar\NAT@@open\if*#1*\else#1\NAT@spacechar\fi%
   \NAT@hyper@{\NAT@date}}
  {}
  {}
\makeatother

\iflocal
\def\figdir{figs}
\else
\def\figdir{.}
\fi

\defcitealias{diemer_15}{DK15}
\defcitealias{diemer_14}{DK14}
\defcitealias{diemer_13_scalingrel}{DKM13}

\shorttitle{Diemer \& Joyce}
\shortauthors{Diemer \& Joyce}

\journalinfo{The Astrophysical Journal, {\rm 871:168 (16pp), 2019 February 1}}
\submitted{Received 2018 September 19; revised 2018 December 18; accepted 2018 December 21; published 2019 January 30}

\begin{document}

\title{An accurate physical model for halo concentrations}
\author{Benedikt Diemer\altaffilmark{1} and Michael Joyce\altaffilmark{1,2}}

\affil{
$^1$ Institute for Theory and Computation, Harvard-Smithsonian Center for Astrophysics, 60 Garden St., Cambridge, MA 02138, USA; \href{mailto:benedikt.diemer@cfa.harvard.edu}{benedikt.diemer@cfa.harvard.edu} \\
$^2$ Laboratoire de Physique Nucl\'eaire et de Hautes \'Energies, UPMC IN2P3 CNRS UMR 7585, Sorbonne Universit\'e, 4, place Jussieu, F-75252 Paris Cedex 05, France
}


\begin{abstract}
The relation between halo mass, $M$, and concentration, $c$, is a critical component in our understanding of the structure of dark matter halos. While numerous models for this relation have been proposed, almost none of them attempt to derive the evolution of the relation analytically. We build on previous efforts to model the \cmr as a function of physical parameters such as the peak height, $\nu$, and the effective power spectrum slope, $\neff$, which capture the dependence of $c$ on halo mass, redshift, and  cosmology. We present three major improvements over previous models. First, we derive an analytical expression for the \cmr that is valid under the assumption of pseudo-evolution, i.e., assuming that the density profiles of halos are static in physical coordinates while the definition of their boundary evolves. We find that this ansatz is highly successful in describing the  evolution of the low-mass end of the \cmr. Second, we employ a new physical variable, the effective exponent of linear growth, $\alphaeff$, to parameterize deviations from an Einstein--de Sitter expansion history. Third, we combine an updated definition of $\neff$ with the additional dependence on $\alphaeff$ and propose a phenomenological extension of our analytical framework to include all halo masses. This semianalytical model matches simulated concentrations in both scale-free models and \LCDM to 5\% accuracy with very few exceptions and differs significantly from all previously proposed models. We present a publicly available code to compute the predictions of our model in the python toolkit \colossus, including updated parameters for the model of Diemer and Kravtsov.
\end{abstract}



\section{Introduction}
\label{sec:intro}

The density structure of dark matter halos is a critical ingredient in modeling observations of galaxies and galaxy clusters, making the spherically averaged halo density profile, $\rho(r)$, a key physical quantity. The density profiles must, of course, depend on the total mass $M$ of a halo, but are they otherwise universal, or does their shape depend on halo mass, redshift, and cosmology? Virtually all forms of the density profile that have been proposed in the literature had to rely on an additional parameter, a scale radius $\rs$ \citep{einasto_65, einasto_69, hernquist_90, navarro_97, navarro_04}, often defined as the radius where the logarithmic slope of the density profile reaches $-2$. In particular \citet[][hereafter NFW]{navarro_95, navarro_96, navarro_97} claimed that density profiles depend only on mass and scale radius \citep[see, however,][]{diemer_14} and proposed a convenient parameterization where the scale radius is expressed as concentration, defined as the ratio of an outer radius to the scale radius, $c = R/\rs$. This additional parameter breaks the universality of the profiles at fixed halo mass unless it can itself be described as a function of mass: the so-called \cmr.

The \cmr was found, however, to exhibit complex dependencies on redshift and cosmology. Numerous proposals for how to model these dependencies have been put forward, most of which fall into two categories. First, NFW suggested that concentration is intimately linked to the age of a halo or, more generally, its assembly history. This idea proved to be a fruitful avenue toward building age-based models that predict both the average concentration and scatter \citep{navarro_96, navarro_97, bullock_01, eke_01, wechsler_02, zhao_03_concentration, lu_06, dalal_08, zhao_09, giocoli_12, ludlow_14, ludlow_16, vandenbosch_14, correa_15_c}. Another popular way to describe the \cmr is to simply fit average concentrations, typically with power laws or other simple functions \citep{avilareese_99, jing_00_profiles2, colin_04, dolag_04, neto_07, duffy_08, gao_08, maccio_08, klypin_11, munozcuartas_11, bhattacharya_13, dutton_14, heitmann_15, klypin_16, hellwing_16, child_18}. Such fitting functions are valid only for the redshift and cosmology where they were constrained \citep[unless they are interpolated as in][]{kwan_13}. Moreover, power-law fits predict demonstrably wrong concentrations when extrapolated to very low halo masses \citep{ludlow_14}.

More recently, a third type of model for the \cmr has emerged. \citet{prada_12} noted that concentrations exhibit a much less drastic redshift evolution if mass is expressed as peak height, $\nu$, the statistical significance of a peak over the linear density field (see Section~\ref{sec:theory:defs} for the exact definition). They parameterized the remaining dependence with an empirical fitting function. \citet[][hereafter \citetalias{diemer_15}]{diemer_15} showed that the deviations from universality can be understood physically by adding a second variable besides peak height: the effective slope of the power spectrum, $\neff$. While similar dependencies on the power spectrum had been considered before \citep{bullock_01, eke_01, zhao_09}, \citetalias{diemer_15} write concentration as a function of only $\nu$ and $\neff$. With a modest seven free parameters that are fitted to simulation data, their function describes the concentrations in both \LCDM cosmologies and scale-free, self-similar Einstein--de Sitter (EdS) universes and over a vast range of masses and redshifts.

While it is encouraging that such different types of models successfully describe the \cmr, most of them share one shortcoming: whatever physical mechanisms shape concentration are perhaps understood in broad strokes but do not directly inform the functional form of the \cm or \cnur (for partial exceptions see \citealt{ludlow_14} and \citealt{okoli_16}). For example, while it is enlightening to understand that concentration increases with halo age, it is not obvious {\it how} concentration evolves as a function of time. Similarly, we know that the power spectrum slope plays a role in controlling concentration, but we can only speculate about the exact mechanisms \citep[][\citetalias{diemer_15}]{navarro_97, eke_01, reed_05_profiles, knollmann_08, zhao_09}. For example, a shallow $\neff$ leads to a shallow mass function, to increased mergers with subhalos that preferably sink to the center owing to dynamical friction, and thus to higher concentration \citep{chandrasekhar_43, lacey_93, boylankolchin_08, rey_18}. Second, the shape of peaks in Gaussian random fields is determined by $\neff$ and is known to affect the final profile shape \citep{bardeen_86, dalal_10}. However, it remains unclear how, quantitatively speaking, these mechanisms manifest themselves in the \cmr.

In this work, we make a significant step in improving this situation by providing an analytical derivation of the \cmr that accurately describes its low-mass end. We then build on this derivation and robust results from scale-free simulations to construct a simple ansatz extending over the full mass range. The fundamental idea of our analytical derivation is that the evolution of halos with low peak height, i.e., halos that formed some time ago, is relatively simple. We can imagine the formation of a halo as a two-stage process: in the early fast-accretion regime, the halo grows rapidly and its profile maintains a roughly universal shape with $\cvir \approx 4$ \citep{zhao_03_mah, zhao_09}. Once this growth slows down, the scale radius of the halo approaches a constant value, meaning that the center of the halo remains more or less static in physical coordinates \citep{bullock_01, ludlow_13}. At this point, the halo starts to grow largely because of ``pseudo-evolution,'' a growth in radius and mass due to the changing reference density used to define the halo boundary \citep{diemand_05, cuesta_08, diemer_13_pe, diemer_13_scalingrel, zemp_14, more_15}. For the purposes of our investigation, it is not relevant whether pseudo-evolution is attributed to physical mass accretion outside of the initial halo radius or to the changing halo boundary, as long as the evolution of the radius is governed by the initial density profile. \citet{diemer_13_pe} showed numerically that the corresponding evolution of concentration reproduces the trends observed in simulations at low masses (their Figure~9).

Motivated by this finding, we attempt to combine two types of \cm modeling: at low masses, we refer to the connection between concentration and halo age by deriving the time evolution in the pseudo-evolving limit exactly, and at high masses, we adopt an approach similar to \citetalias{diemer_15} in that we extend our low-mass model phenomenologically, explaining any nonuniversality of the \cnur with physical parameters such as $\neff$. We find that pseudo-evolution is, indeed, an excellent description of the \cnur at $\nu \lsim 1.4$, and we provide few-parameter fitting functions for this regime. Motivated by seeming discrepancies between scale-free and \LCDM simulation data at fixed $\nu$ and $\neff$, we add a third variable, the effective exponent of linear growth, $\alphaeff$. This physical extension improves our fit systematically compared to \citetalias{diemer_15} despite needing one fewer free parameter (six instead of seven).

The paper is structured as follows. In Section~\ref{sec:sims}, we briefly describe the simulation data used in this paper, referring the reader to \citetalias{diemer_15} for details. We derive our semianalytical expression for the \cmr in Section~\ref{sec:theory} and compare its predictions to previous works in Section~\ref{sec:comparison}. We further discuss our results in Section~\ref{sec:discussion} and summarize our conclusions in Section~\ref{sec:conclusions}. Finally, in Appendix~\ref{sec:app:dk15update}, we provide an updated version of the \citetalias{diemer_15} model that corrects a small numerical error in the original analysis. We consider the conversion between mass definitions in Appendix~\ref{sec:app:mdef}.


\begin{deluxetable*}{lccccccccl}
\tablecaption{$N$-body Simulations
\label{table:sims}}
\tablewidth{0pt}
\tablehead{
\colhead{Name} &
\colhead{$L\, (\mpch)$} &
\colhead{$N^3$} &
\colhead{$m_{\rm p}\, (\msunh)$} &
\colhead{$\epsilon\, (\kpch)$} &
\colhead{$\epsilon / (L / N)$} &
\colhead{$z_{\rm initial}$} &
\colhead{$z_{\rm final}$} &
\colhead{Cosmology} &
\colhead{Reference}
}
\startdata
L2000 & $2000$ & $1024^3$ & $5.6 \times 10^{11}$  & $65$  & $1/30$ & $49$ & $0$ & $WMAP$ (Bolshoi) & \citetalias{diemer_15} \\
L1000 & $1000$ & $1024^3$ & $7.0 \times 10^{10}$ & $33$ & $1/30$ & $49$ & $0$ & $WMAP$ (Bolshoi) & \citetalias{diemer_13_scalingrel} \\
L0500 & $500$  & $1024^3$ & $8.7 \times 10^{9}$  & $14$ & $1/35$  & $49$ & $0$ & $WMAP$ (Bolshoi) & \citetalias{diemer_14} \\
L0250 & $250$  & $1024^3$ & $1.1 \times 10^{9}$  & $5.8$  & $1/42$  & $49$ & $0$ & $WMAP$ (Bolshoi) & \citetalias{diemer_14} \\
L0125 & $125$  & $1024^3$ & $1.4 \times 10^{8}$  & $2.4$  & $1/51$  & $49$ & $0$ & $WMAP$ (Bolshoi) & \citetalias{diemer_14} \\
L0063 & $62.5$ & $1024^3$ & $1.7 \times 10^{7}$  & $1.0$  & $1/60$ & $49$ & $0$ & $WMAP$ (Bolshoi) & \citetalias{diemer_14} \\
L0031 & $31.25$ & $1024^3$ & $2.1 \times 10^{6}$  & $0.25$  & $1/122$ & $49$ & $2$ & $WMAP$ (Bolshoi) & \citetalias{diemer_15} \\
L0500-Planck & $500$  & $1024^3$ & $1.0 \times 10^{10}$  & $14$ & $1/35$  & $49$ & $0$ & $Planck$ & \citetalias{diemer_15} \\
L0250-Planck & $250$  & $1024^3$ & $1.3 \times 10^{9}$  & $5.8$  & $1/42$  & $49$ & $0$ & $Planck$ & \citetalias{diemer_15} \\
L0125-Planck & $125$  & $1024^3$ & $1.6 \times 10^{8}$  & $2.4$  & $1/51$  & $49$ & $0$ & $Planck$ & \citetalias{diemer_15} \\
L0100-PL-1.0 & $100$  & $1024^3$ & $2.6 \times 10^{8}$  & $0.5$  & $1/195$  & $119$ & $2$ & Self-similar, $n=-1.0$ & \citetalias{diemer_15} \\
L0100-PL-1.5 & $100$  & $1024^3$ & $2.6 \times 10^{8}$  & $0.5$  & $1/195$  & $99$ & $1$ & Self-similar, $n=-1.5$ & \citetalias{diemer_15} \\
L0100-PL-2.0 & $100$  & $1024^3$ & $2.6 \times 10^{8}$  & $1.0$  & $1/98$  & $49$ & $0.5$ & Self-similar, $n=-2.0$ & \citetalias{diemer_15} \\
L0100-PL-2.5 & $100$  & $1024^3$ & $2.6 \times 10^{8}$  & $1.0$  & $1/98$  & $49$ & $0$ & Self-similar, $n=-2.5$ & \citetalias{diemer_15}
\enddata
\tablecomments{The $N$--body simulations used in this paper. $L$ denotes the box size in comoving units, $N^3$ the number of particles, $m_{\rm p}$ the particle mass, and $\epsilon$ the force softening length in physical units. The references correspond to \citet[][\citetalias{diemer_13_scalingrel}]{diemer_13_scalingrel}, \citet[][\citetalias{diemer_14}]{diemer_14}, and \citet[][\citetalias{diemer_15}]{diemer_15}. Our system for choosing force resolutions is discussed in \citetalias{diemer_14}.}
\end{deluxetable*}

\vspace{0.5cm}

\section{Simulation Data}
\label{sec:sims}
 
In this section, we describe our simulation data, halo samples, and algorithms for fitting and binning concentrations.

\subsection{$N$-body Simulations}
\label{sec:sims:sims}

We use essentially the same suite of dissipationless $N$-body simulations as \citetalias{diemer_15}. These simulations include scale-free EdS and \LCDM cosmologies in different box sizes; their detailed properties are listed in Table~\ref{table:sims}. The EdS simulations have power spectrum slopes of $-1$, $-1.5$, $-2$, and $-2.5$. The \LCDM simulations use two different, flat cosmologies. The first cosmology is that of the {\it Bolshoi} simulation \citep{klypin_11}, which is consistent with {\it WMAP7} \citep[][$\Omega_{\rm m} = 0.27$, $\Omega_{\rm b} = 0.0469$, $h = 0.7$, $\sigma_8 = 0.82$, and $n_{\rm s} = 0.95$]{komatsu_11}. For this cosmology, we use seven boxes with side lengths decreasing by factors of two from $2000$ down to $31.25 \mpch$. The second cosmology is similar to the \citet{planck_14} cosmology ($\Omega_{\rm m} = 0.32$, $\Omega_{\rm b} = 0.0491$, $h = 0.67$, $\sigma_8 = 0.834$, and $n_{\rm s} = 0.9624$). For this cosmology, we use three boxes of $500$, $250$, and $125 \mpch$, respectively. The two cosmologies bracket the currently favored range of possible cosmological parameters.

The initial conditions for the simulations were generated using a \textsc{Camb} power spectrum \citep{lewis_00} and the \textsc{2LPTic} code \citep{crocce_06}, and the simulations were evolved with \textsc{Gadget2} \citep{springel_05_gadget2}. We use the \textsc{Rockstar} and \textsc{Consistent-Trees} codes \citep{behroozi_13_rockstar, behroozi_13_trees} to construct halo catalogs and merger trees. \textsc{Rockstar} finds the particles in friends-of-friends groups in six-dimensional phase space and uses the gravitationally bound particles to compute the properties of halos. Our catalogs are based on $\rvir$ as the halo radius (Section~\ref{sec:theory:defs}).

\subsection{Halo Selection}
\label{sec:sims:sample}

\citetalias{diemer_15} determined the minimum mass for halos at a given redshift as the maximum of three resolution criteria. In particular, they required \num{1000} particles inside $\rtoc$, $200$ particles inside $\rs$, and that $\rs$ be at least six times larger than the force resolution of a given simulation. If these requirements were enforced for individual halos, they would lead to a biased scale radii. Instead, they are enforced on average, i.e., by computing a halo mass $\mvir$ that fulfills the requirements assuming the \cmr of \citet{zhao_09}. Due to an erroneous conversion from physical to comoving units, the final criterion was too strict in \citetalias{diemer_15}, which is why the \cm data used in this work extend to slightly higher masses at high redshift. 

We make no cuts on the dynamical state of our halos, including even unrelaxed systems that may be poorly fit by NFW profiles. In doing so, we accept that concentration may be ill-defined for some halos, contributing to tails of extreme values in the distribution (e.g., Figure~1 in \citetalias{diemer_15}). Unrelaxed systems have been shown to cause the upturn in our \cnu relations at high peak height. On the other hand, removing the upturn by aggressively cutting out unrelaxed halos necessarily leads to a bias toward dynamically older systems and thus higher concentrations \citep{ludlow_12, prada_12, correa_15_c, angel_15, klypin_16, child_18}. To avoid such biases, we consider the full halo sample.

All halos that pass the three resolution criteria are combined into one sample per redshift per cosmology, regardless of which simulation they originated from. For the scale-free models, the evolution of the \cm relation represents merely a rescaling of mass (as we show in Section~\ref{sec:theory:pe_eds}), and we thus combine halos from multiple redshifts into one sample per cosmology. We refer the reader to \citetalias{diemer_15} for further details on these halo samples. 

\subsection{Fitting and Binning}
\label{sec:sims:fitting}

We rely on the concentrations measured by the \textsc{Rockstar} halo finder, which fits an NFW profile to the mass profile of each halo. For this purpose, \textsc{Rockstar} groups all bound particles within $\rvir$ into up to $50$ equal-mass bins with at least $15$ particles per bin. The scale radius is varied until the best-fit NFW profile has been found. All bins receive equal weight except for bins that lie within three force resolution lengths, which are down-weighted by a factor of $10$ \citep{behroozi_13_rockstar}.

We note that there are alternative ways to measure concentration. For example, \citet{klypin_11} suggested using the circular velocities at certain radii instead of fitting an NFW profile, leading to systematically different values. Concentrations can also be derived from Einasto profiles, which tend to provide a better fit \citep{navarro_04}, but this profile form relies on an additional shape parameter, $\alpha$. This shape parameter needs to be fixed when fitting to individual halos because their profiles often allow too much freedom, but $\alpha$ depends on peak height and the shape of the power spectrum in a nontrivial fashion \citep{gao_08, ludlow_17}. \citet{dutton_14} showed that Einasto and NFW concentrations agree to about 10\% or better (their Figure~5; see also \citealt{meneghetti_14}). We have checked that this conclusion applies to both scale-free and $\Lambda$CDM universes and that those models do not systematically differ in the relative fit quality of NFW and Einasto profiles. Thus, we refrain from exploring different definitions of concentrations and rely on the NFW fits performed by \textsc{Rockstar}.

The concentrations of all halos that pass the resolution cuts are binned in logarithmic mass or peak height. The shaded error regions shown in the forthcoming figures correspond to the statistical uncertainty on the mean or median, not the scatter, which is much larger, about $0.16$ dex \citepalias{diemer_15}. 

All fits are performed using the binned concentration data (rather than the concentrations of individual halos) and use a standard least-squares algorithm. The best-fit parameters are often largely determined by low-mass bins that contain many more halos than their high-mass counterparts. However, even if the median value in a bin is statistically well determined, it carries a systematic uncertainty due to profile fitting, binning, and numerical effects. Thus, we add a systematic error to 2\% of the concentration and compute the uncertainty on a bin's value as $\sigma_{\rm bin} = \sqrt{\sigma_{\rm stat}^2 + (0.02 c_{\rm bin})^2}$.

Due to a numerical error in \citetalias{diemer_15}, their concentrations for the {\it WMAP7} cosmology were underestimated at all redshifts except $z = 0$ by up to 5\%. All conclusions of their work remain valid, and their figures change relatively little. We have, of course, corrected this error, and we provide an updated version of the \citetalias{diemer_15} best-fit parameters in Appendix~\ref{sec:app:dk15update}.


\section{Semianalytical Models of the \cm Relation}
\label{sec:theory}

In this section, we derive a semianalytical description of the evolution of halo concentration. We begin by defining a number of variables (Section~\ref{sec:theory:defs}) and considering the pseudo-evolution of concentration in general (Section~\ref{sec:theory:pe}). We then break up the task of finding an analytical expression for the \cm relation by considering four levels of complexity: low-mass halos in scale-free cosmologies (Section~\ref{sec:theory:pe_eds}), low-mass halos in arbitrary \LCDM cosmologies (Section~\ref{sec:theory:pe_lcdm}), all halos in scale-free cosmologies (Section~\ref{sec:theory:all_eds}), and finally all halos in \LCDM (Section~\ref{sec:theory:all_lcdm}). Readers who wish to skip the mathematical details of our derivation may proceed to Section~\ref{sec:theory:all_lcdm} where we demonstrate the quality of our fitting function (Equation~(\ref{equ:fit_func}), Figure~\ref{fig:main_fit}).

\subsection{Definitions and General Considerations}
\label{sec:theory:defs}

We use $r$ to denote three-dimensional radii measured from the halo center and reserve capital letters such as $R$ and $M$ for specific radii and masses used to define the halo boundary. In particular, spherical overdensity (SO) radii are defined as
\begin{equation}
\rdelta = \left(\frac{3\mdelta}{4\pi \Delta \rhoref(z)}\right)^{1/3} \,,
\label{equ:mass_so}
\end{equation}
where $\rho_{\rm ref}$ is either the critical or mean density of the universe and $\Delta$ is a dimensionless overdensity, leading to definitions such as $\rtoc$ or $\rtom$ (where $\Delta$ is set to $200$ times the critical and mean density of the universe, respectively). We use the subscript ``vir'' to indicate quantities that are calculated based on a varying overdensity $\Delta_{\rm vir}(z)$, computed using the approximation of \citet{bryan_98}. For much of this work, we express halo mass as peak height, $\nu$, which is defined as
\begin{equation}
\nu_{\Delta} \equiv \frac{\delta_{\rm c}}{\sigma(\mdelta, z)} = \frac{\delta_{\rm c}}{\sigma(\mdelta, z = 0) \times D(z)} \,.
\label{eq:nu}
\end{equation}
Here $\delta_{\rm c} = 1.686$ denotes the critical overdensity for top-hat collapse in an EdS universe \citep{gunn_72}, $D(z)$ is the linear growth factor of density fluctuations normalized to unity at $z = 0$ \citep[e.g.][]{eisenstein_99}, and $\sigma(M)$ denotes the rms density fluctuation in spheres of the corresponding Lagrangian radius. This radius is defined with respect to some $\mdelta$ such that 
\begin{equation}
M_{\rm L} = \mdelta = (4 \pi/3) \rho_{\rm m}(z=0) R_{\rm L}^3 \,.
\end{equation}
For much of the paper, we will use $\nutoc$, whereas the letter $\nu$ indicates a generalized form that does not depend on the exact mass definition. The nonlinear mass $M_*$ is defined as the mass where $\nu = 1$. We use the fitting function of \citet{eisenstein_98} to compute the linear power spectrum on which the variance is based. In principle, we could use the exact power spectra from the \textsc{Camb} code, which was used to create the initial conditions of our simulations. In practice, however, using an analytical approximation makes the calculation of our model much faster and more portable. All calculations are performed using the python toolkit \textsc{Colossus} \citep{diemer_18_colossus}.

In this work, we are concerned with concentration, defined as $\cdelta = \rdelta / \rs$. Here $\rs$ is the scale radius where the logarithmic slope of the profile is $-2$. By similarly defining a dimensionless radius variable $x = r/\rs$ and a scale density $\rhos$, we can write any density profile as the enclosed mass,
\begin{equation}
M(<r) = 4 \pi \rhos \rs^3 g(x) \,,
\label{equ:mass_rs}
\end{equation}
where $g(x)$ is an arbitrary function that depends on the density profile. For example, for the NFW profile,
\begin{equation}
\rho(r) = \frac{\rhos \rs^3}{r(r + \rs)^2} \,,
\end{equation}
we have
\begin{equation}
g(x) = \ln (1 + x) - \frac{x}{1 + x} \,.
\label{equ:gc-nfw}
\end{equation}
We will use the NFW profile when evaluating our expressions for concentration, but we emphasize that those expressions are general and should hold for any $g(x)$.

\subsection{Pseudo-evolving Halo Concentrations}
\label{sec:theory:pe}

As discussed in Section~\ref{sec:intro}, we have reason to suspect that the concentrations of low-mass halos change mostly as a result of pseudo-evolution. Here the term ``pseudo-evolution'' means that the scale radius and central density profile of a halo are assumed to be constant in physical coordinates, and thus that they evolve solely through the growth of their outer boundary $\rdelta$. \citet{diemer_13_pe} demonstrated that this assumption can reproduce the evolution of the low-mass end of the \cmr according to the accretion history model of \citet{zhao_09}. This finding highlights that pseudo-evolution is implicit in some models of halo growth, where $\rs$ stays constant but SO radii evolve.

We note that our definition of the term ``pseudo-evolution'' is less strict than in other contexts where it is sometimes used to describe the gradual inclusion of mass that had already been physically accreted onto the halo at the initial time. We remain agnostic about whether the halo mass grows because matter is being accreted after the initial time or because of the artificial (``pseudo'') mass growth caused by the arbitrary definition of the halo boundary.

To build an analytical model of pseudo-evolving concentrations, let us now assume that, at any redshift $z$, all halos of mass less than some mass $\mpe(z)$ are pseudo-evolving halos and that each such halo enters this phase at a redshift $\zpe \geq z$ depending on $\mpe$. We denote by $\cpe$ the concentration of a halo at the redshift $\zpe$ when it entered the pseudo-evolving regime. As $\rhos$ and $\rs^3$ are then constant in physical coordinates, it follows that $\mdelta \propto g(c)$. Equations (\ref{equ:mass_so}) and (\ref{equ:mass_rs}) then give the following two equations:
\begin{equation}
\frac{c}{ \left[g(c)\right]^{\frac{1}{3}}}=\left(\frac{\Delta \rhoref (\zpe) (1+z)^3}{\Delta \rhoref (z) (1+\zpe)^3}\right)^{\frac{1}{3}} \frac{1+\zpe}{1+z} \frac{\cpe}{ \left[g(\cpe)\right]^{\frac{1}{3}}}
\label{equ:c_pe_general}
\end{equation}
and 
\begin{equation}
\frac{\mdelta}{g(c)}= \frac{\mpe}{g(\cpe)} \,.
\label{equ:c_pe_general2}
\end{equation}
Defining
\begin{equation}
F(x)=\frac{x}{ \left[g(x)\right]^{\frac{1}{3}}}
\end{equation}
and denoting the first factor on the right-hand side of Equation (\ref{equ:c_pe_general}) as $X_{\rm ref} (z)$, we can conveniently write Equation (\ref{equ:c_pe_general}) as 
\begin{equation}
c = \tilde{F} \left[ X_{\rm ref} (z) \frac{1+\zpe}{1+z} F(\cpe) \right] \,,
\label{equ:c_pe_convenient}
\end{equation}
where $\tilde{F} $ is the inverse function of $F(x)$ (which is a monotonically increasing function defined for $x \geq 0$). We note that this equation is valid for all mass definitions, with $X_{\rm ref} (z) = 1$ when $\rhoref = \rhom$ and $X_{\rm ref}(z) = \Omega_{\rm m}(z)/\Omega_{\rm m}(\zpe)$ if $\rhoref = \rhoc$. At any $z$, Equations (\ref{equ:c_pe_general2}) and (\ref{equ:c_pe_convenient}) allow us to calculate, for any $\mpe$, the values of $c$ and $\mdelta$, provided we know $\zpe$ and $\cpe$ as a function of $\mpe$. 

\subsection{Low-mass Halos in Scale-free Cosmologies}
\label{sec:theory:pe_eds}

\begin{figure*}
\centering
\includegraphics[trim =   7mm 10mm   2mm 0mm, clip, scale=0.65]{\figdir/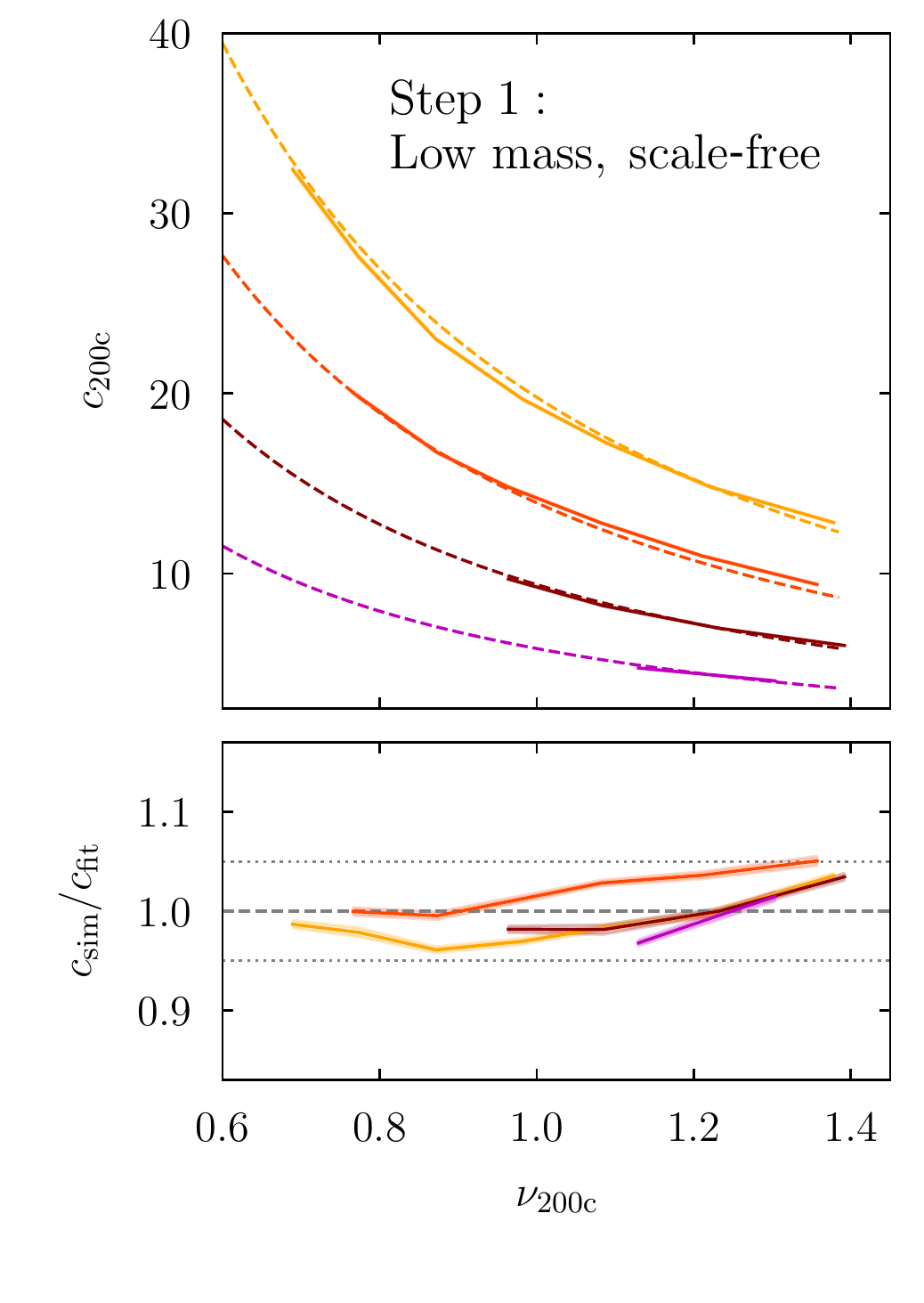}
\includegraphics[trim =  15mm 10mm 164mm 0mm, clip, scale=0.65]{\figdir/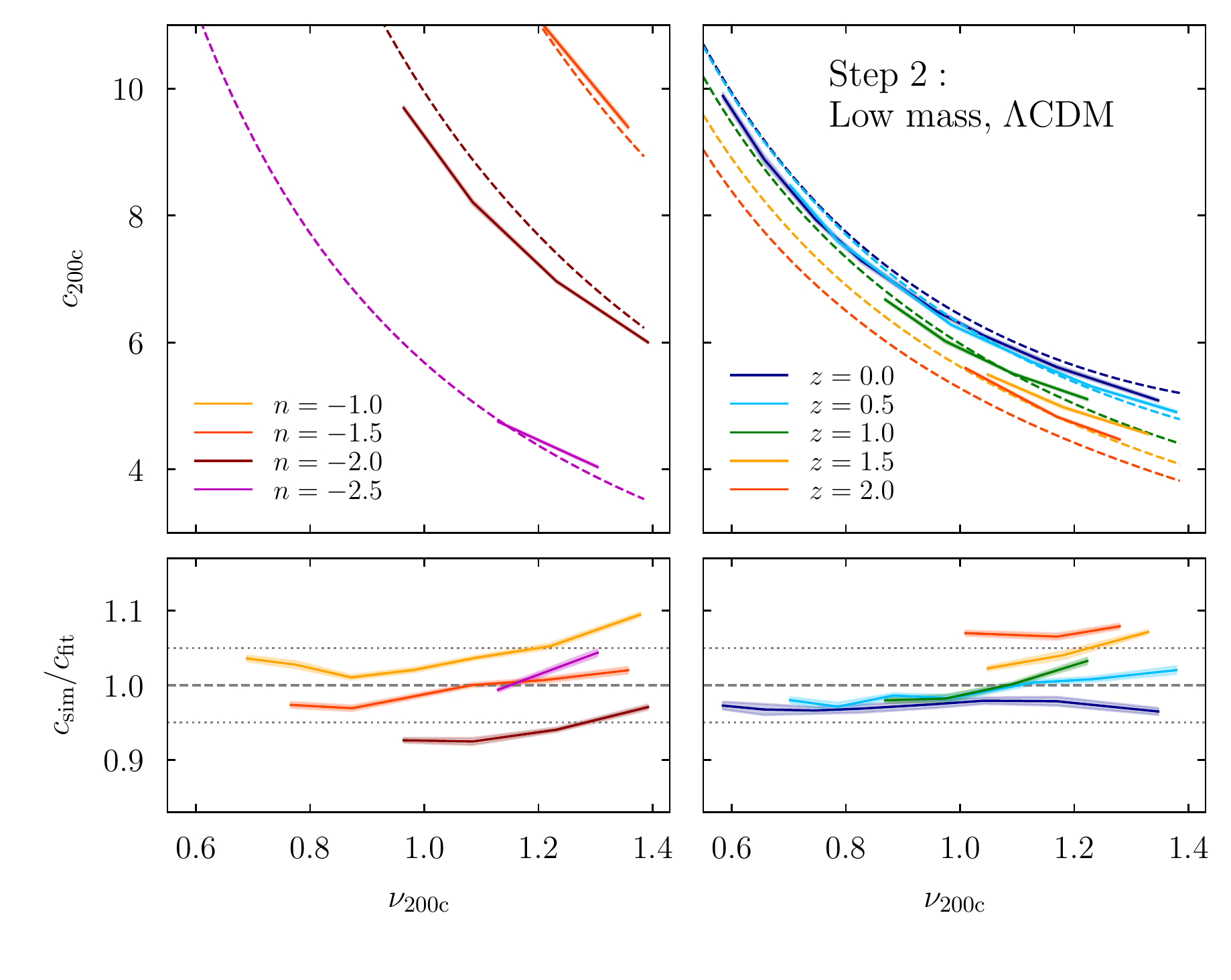}
\includegraphics[trim = 105mm 10mm   2mm 0mm, clip, scale=0.65]{\figdir/fit_step_2.pdf}
\includegraphics[trim =  15mm 10mm   2mm 0mm, clip, scale=0.65]{\figdir/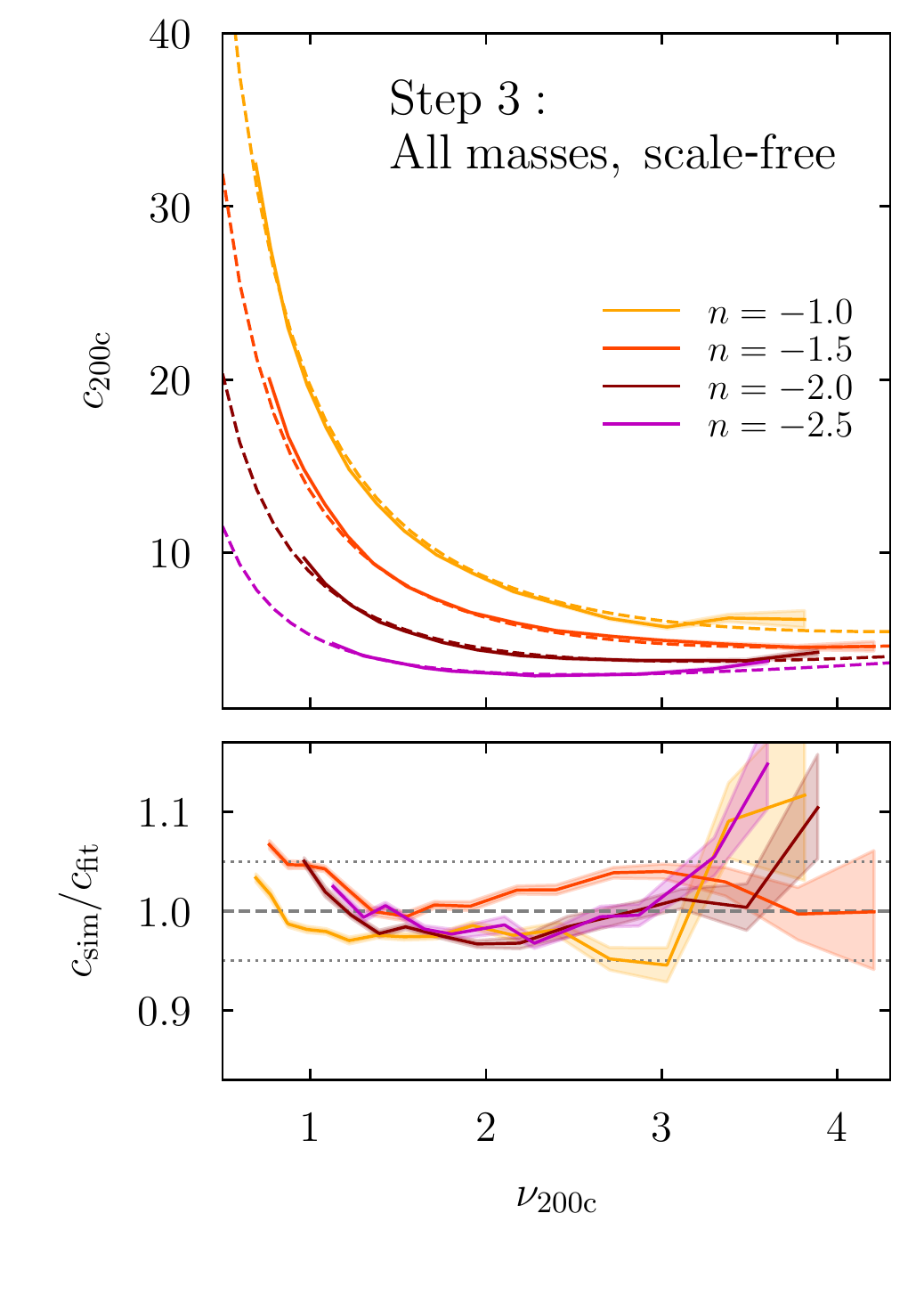}
\caption{Intermediate steps in the development of our semianalytical model. Each set of panels shows the median concentration (solid lines), statistical uncertainty (shaded area), and fit (dashed lines) for a given set of \cnu data. The bottom panels show the relative difference between fit and data. {\it Left panels:} A two-parameter fit to the low-mass (pseudo-evolving) end of the \cnur in scale-free models (Equation~\ref{equ:fit_step1}, Section~\ref{sec:theory:pe_eds}). {\it Middle panels:} A three-parameter fit to the low-mass end for both scale-free (not shown) and \LCDM cosmologies (Section~\ref{sec:theory:pe_lcdm}). {\it Right panels:} A four-parameter fit to all halo masses in scale-free models (Equation~\ref{equ:fit_step3}, Section~\ref{sec:theory:all_eds}). See the respective sections for a detailed discussion.}
\label{fig:fit_steps}
\end{figure*}

The functions above take on particularly simple forms in scale-free cosmologies, and our goal is to derive their form in more general \LCDM cosmologies by an interpolation of such models. By ``scale-free models'' we mean CDM cosmologies with an initial linear power spectrum that is a simple power law, $P(k) \propto k^n$ with $n$ constant, and an EdS expansion law, $a \propto t^{2/3}$. Because of the absence of characteristic mass or length scales, clustering in these models must be self-similar, i.e., its temporal evolution must be equivalent to a rescaling of lengths or masses. In the present context, this property implies that the time dependence of the \cmr can be removed by rescaling masses:
\begin{equation}
c(M,z) = c_0 \left( \frac{M}{M_*(z)} \right) \,,
\label{cm-scalefree}
\end{equation}
where $M_*(z)$ is the characteristic mass scale defined at redshift $z$ (defined relative to some arbitrary reference time at which $z = 0$). In the scale-free case, $\sigma^2 (M,z) \propto  M^{-\frac{n+3}{3}}$, and therefore 
\begin{equation}
\sigma(M, z) = \delta_c \left(\frac{M_*}{M}\right)^{\frac{n+3}{6}} \,.
\end{equation}
Given that the linear growth in an EdS cosmology is proportional to the scale factor, $\sigma(M, z) \propto 1/(1+z)$, we infer 
\begin{equation}
M_* (z) \propto \left(\frac{1}{1+z}\right)^\frac{6}{3+n} \,.
\label{mstar-a}
\end{equation}   
The peak height is then $\nu(M,z) = \nu (M,0) (1+z)$ and thus
\begin{equation}
\nu(M,z)= \left(\frac{M}{M_*(z)}\right)^{\frac{n+3}{6}} \,.
\label{nu-m-relation-SF}
\end{equation}
This self-similarity implies that $c(\nu)$ is a time-independent function in a scale-free model (which is borne out in simulations; \citetalias{diemer_15}). It is now straightforward to find the functional form of the \cm relation in the case of pure pseudo-evolution. Since $X_{\rm ref} = 1$, we can write Equation (\ref{equ:c_pe_convenient}) as 
\begin{equation}
\frac{c}{g(c)^{1/3}}=\frac{\nupe}{\nu(\mpe,z)} \frac{\cpe}{g(\cpe)^{1/3}}
\label{relations-pe-sf}
\end{equation} 
where $\nupe = \nu(\mpe,\zpe)$. Combining this expression with Equations (\ref{equ:c_pe_general2}) and (\ref{nu-m-relation-SF}), we obtain
\begin{equation}
\frac{c}{\left[g(c)\right]^{\frac{5+n}{6}}} = \frac{A(n)}{ \nu} \,,
\label{equ:fit_step1}
\end{equation}
where $A(n) = \nupe \cpe / g(\cpe)^{\frac{5+n}{6}}$ depends only on $n$ because $\nupe$ and $\cpe$ are fixed parameters for a given scale-free model. We have thus obtained an implicit analytical expression for the \cnu relation for any $\Delta$, given a density profile $g(c)$. For a given $n$, there is a single free parameter that fixes the overall amplitude. Note that we have assumed, as appropriate in an EdS universe, that $\Delta \rhoref \propto a^{-3}$. The dependence of $\nu$ on $\Delta$ is then absorbed in the constant $A(n)$. Thus, the functional form of the pseudo-evolving \cmr in a scale-free cosmology is independent of the detailed definition of the halo boundary.

We solve Equation~(\ref{equ:fit_step1}) numerically to obtain $c(\nu, n)$. We expect this expression to work only for low halo masses where pseudo-evolution dominates \citep{diemer_13_pe}. By experimenting, we find that $\nucut = 1.4$ appears to capture the transition to the regime where the median concentration begins to deviate from Equation~(\ref{equ:fit_step1}) by more than about 10\%. There is no evidence that this limit depends on $n$. Leaving $A(n)$ free for each scale-free model, we obtain excellent fits to the \cm data from our scale-free simulations. When fitting all four scale-free models ($n = [-1, -1.5, -2, -2.5]$) simultaneously with $\nupe$ and $\cpe$ as free parameters, we obtain a fit quality of about 10\%. However, $\nupe$ and $\cpe$ are degenerate, and we find a better fit when assuming that $A(n)$ has a linear relation with $n$, 
\begin{equation}
A(n) = a_0 + a_1 (n + 3) \,.
\end{equation}
Writing the slope as $(n + 3)$ is not necessary but leads to more intuitive values of the best-fit parameters. Fitting the $\nu < 1.4$ scale-free simulation data for $\ctoc$, we match the data to better than 5\% accuracy with $a_0 = 3.49$ and $a_1 = 4.33$ (Figure~\ref{fig:fit_steps}). As expected, Equation~(\ref{equ:fit_step1}) increasingly underestimates the simulated concentrations toward higher $\nu$ because those halos have gained more mass compared to the expectation from pure pseudo-evolution.

We note that the term on the left-hand side of Equation~(\ref{equ:fit_step1}) plays a crucial role in that it induces an $n$-dependence in the functional form of the \cnu relation. Moreover, the $g(c)$ term forces the relation to deviate from a strict power law, leading to a varying logarithmic slope between about $1.2$ and $1.4$ over the range of $n$ probed by our simulations. These values naturally explain the best-fit low-mass slope in the \citetalias{diemer_15} model. Independently of $n$, the function asymptotically approaches $c \propto 1/\nu$ as $c \rightarrow \infty$, corresponding to the limit in which the halo density profiles are strictly stationary in physical coordinates \citep[the so-called stable clustering limit;][]{peebles_74, davis_77, smith_03_powerspec}. For integrable mass profiles such as the Einasto form, this behavior will be attained at smaller values of $c$ compared to the NFW case. 

\subsection{Low-mass Halos in \LCDM Cosmologies}
\label{sec:theory:pe_lcdm}

\begin{figure}
\centering
\includegraphics[trim = 7mm 0mm 2mm 4mm, clip, scale=0.78]{\figdir/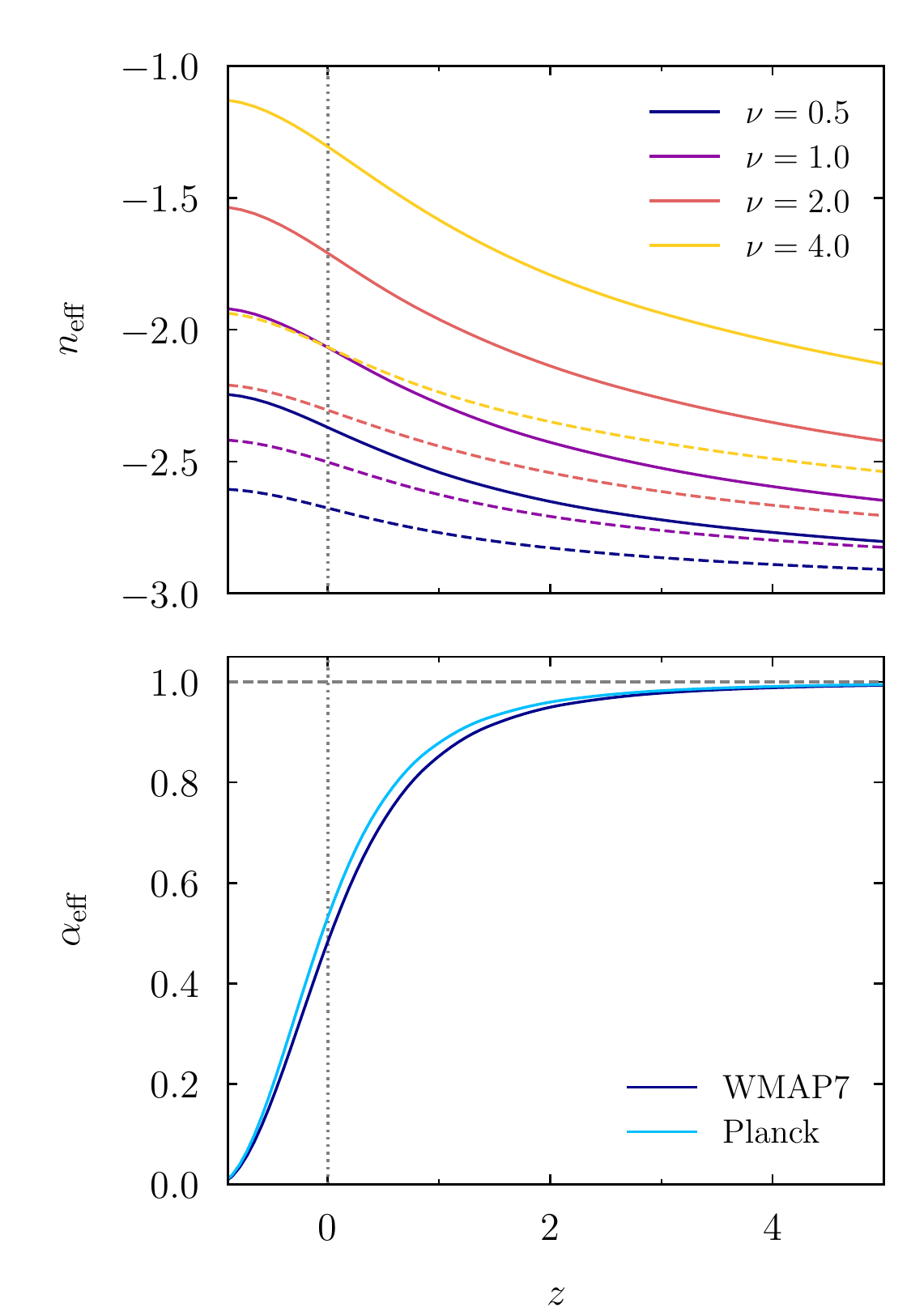}
\caption{Effective slope of the power spectrum, $\neff$ (top panel), and expansion rate, $\alphaeff$ (bottom panel), as a function of redshift. The vertical dotted lines highlight $z = 0$. {\it Top panel:} the effective slope is shown for the {\it WMAP7} cosmology. The solid lines represent the calculation based on the variance, $\sigma(M)$, (Equation~(\ref{neff-1})), and the dashed lines represent the $d \ln P / d \ln k$ version (Equation~(\ref{neff-2})), both with $\kappa = 1$. The $d \ln P / d \ln k$ definition was computed based on the zero-baryon version of the \citet{eisenstein_98} power spectrum to avoid wiggles due to the baryon acoustic oscillations. While both definitions lead to the same overall trend that larger halos experience shallower power spectrum slopes, the dependence on $\nu$ differs in detail. {\it Bottom panel:} the effective expansion rate is defined as $\alphaeff = d \ln (D) / d \ln(1+z)$. Until $z \approx 2$, the expansion is similar to the EdS case of $\alphaeff = 1$, while the growth of structure slows down at low redshift and almost entirely stalls in the future (by $z \approx -0.5$). The differences between the \LCDM cosmologies we consider are small.}
\label{fig:neff_alphaeff}
\end{figure}

We return to Equations~(\ref{equ:c_pe_general2}) and (\ref{equ:c_pe_convenient}), which are general and can be solved numerically for any cosmological model. However, in order to do so, we need to know $\zpe$ and $\cpe$ as a function of $\mpe$, i.e. the redshift at which a halo starts pseudo-evolving and its concentration at that time, when its mass is $\mpe$. As $\zpe$ can be written as
\begin{equation}
\zpe=\tilde{D} \left[ \frac{D(z) \nu(\mpe,z)}{\nupe} \right] \,,
\label{equ:zpe}
\end{equation}
we can phrase the problem as needing to know $\nupe$ and $\cpe$ as a function of $\mpe$. The concentration data from the scale-free models indicate that $\nucut$ is roughly constant, meaning that we can assume that $\nupe$ is constant and thus need only a prescription of $\cpe$ as a function of $\mpe$. Once again, we assume a linear dependence on $\neff$,
\begin{equation}
\cpe = c_0 + c_1 (\neff(\mpe, z) + 3) \,.
\end{equation}
However, the meaning of $n$ is no longer uniquely defined because the slope of the power spectrum is a function of scale in \LCDM cosmologies. 
Thus, we have assumed $n = \neff(M)$, an effective slope that is a function of halo mass. Given the physical meaning of $\cpe$, 
$\neff$ is expected to correspond to the exponent of the scale-free model that best approximates the \LCDM model in the time 
between the formation of a halo and the onset of pseudo-evolution. As we do not know, a priori, how to calculate $\neff$, we consider two plausible prescriptions. First, we could take the logarithmic slope of $\sigma(R)$ at some multiple of the Lagrangian radius of a halo,
\begin{equation}
\neff(M) = -2 \left. \frac{d\ln \sigma(R)}{d \ln R} \right \vert_{R = \kappa R_{\rm L}}-3 \,,
\label{neff-1}
\end{equation}
where $\kappa$ is a free parameter. Second, we could follow \citetalias{diemer_15} and take the logarithmic slope of the power spectrum itself \citep[e.g.,][]{jing_98},
\begin{equation}
\neff (M) = \left. \frac{d \ln P(k)}{d \ln k} \right \vert_{k = \kappa 2 \pi / R_{\rm L}} \,.
\label{neff-2}
\end{equation}
Again, $\kappa$ is a free parameter that was found to be of order unity for the best fits of \citetalias{diemer_15}. The top panel of Figure~\ref{fig:neff_alphaeff} compares these two prescriptions for $\neff$ as a function of redshift and peak height. They evolve similarly with redshift, but their dependence on $\nu$ is different. Finally, we could ignore the mass dependence of $\neff$, postulating that $\cpe$ depends only on redshift, and evaluate Equation~(\ref{neff-1}) at the Lagrangian radius of the nonlinear mass, $R_{\rm L}(M_*)$. This option would correspond to the $\nu = 1$ lines in Figure~\ref{fig:neff_alphaeff}. We find that Equation~(\ref{neff-1}) with a mass-dependent $R_{\rm L}$ leads to the best-fit results and thus adopt it as our prescription for $\neff$.

Combining the implicit Equations~(\ref{equ:c_pe_general2}), (\ref{equ:c_pe_convenient}), and (\ref{equ:zpe}), we numerically solve for the three unknown variables $\mpe$, $\zpe$, and $c$. We fix $\nupe = \nucut = 1.4$, leaving three free parameters ($c_0$, $c_1$, and $\kappa$). We have experimented with a varying $\nupe$ but find that its best-fit value is very close to $\nu_{\rm cut}$. We constrain the parameters in a simultaneous fit to our {\it WMAP7} cosmology and the scale-free models, finding best-fit values of $\kappa = 0.34$, $c_0 = 0.76$, and $c_1 = 5.37$. The middle panels of Figure~\ref{fig:fit_steps} demonstrate that the fitting function with these parameters matches the \LCDM data and scale-free models (not shown) to better than 10\%, though with a discernible redshift trend in the residuals. While we show the $\ctoc$ results, we obtain fits of similar quality when fitting $\ctom$ and $\cfoc$. The equations automatically account for mass definition and can thus be fit to multiple definitions simultaneously (though the fit is slightly degraded in practice).

In summary, we have created a function with only three free parameters that describes the evolution of concentration at low masses in both scale-free and \LCDM cosmologies and for a range of mass definitions. This success demonstrates that 
the \cm relation can be understood almost solely from pseudo-evolution in this regime.

\subsection{All Halos in Scale-free Cosmologies}
\label{sec:theory:all_eds}

Having succeeded at describing the low-mass end of the \cmr with semianalytical functions, we now attempt to expand our understanding to the entire mass range. The physics shaping the \cmr at the high-mass end is complicated. First, we cannot rely on approximations such as pseudo-evolution because halos are physically accreting at rates that, on average, depend on the given cosmology, halo mass, and redshift in a nontrivial fashion. Second, the upturn at the highest peak heights is caused by unrelaxed halos whose density profiles are not well described by an NFW profile \citep{ludlow_12, meneghetti_13}. There is no apparent way to model these trends analytically, which is why we will rely on phenomenological extensions of our low-mass expressions.

The fundamental equations of spherical overdensity radii and the density profile, Equations (\ref{equ:mass_so}) and (\ref{equ:mass_rs}), can be generalized without any assumption about the evolution of halos,
\begin{equation}
\frac{c}{ \left[g(c)\right]^{\frac{1}{3}}}= X_{\rm ref}(z) \frac{1+\zpe}{1+z} \frac{\cpe}{ \left[g(\cpe)\right]^{\frac{1}{3}}}
\prho(z;\mpe,\zpe)
\label{relation1-gen}
\end{equation}
and 
\begin{equation}
\frac{M}{g(c)}= \frac{\mpe}{g(\cpe)} \pmass(z;\mpe,\zpe) \,.
\end{equation}
The functions 
\begin{equation}
\prho (z;\mpe,\zpe)=\left(\frac{\rhos(z)}{\rhos(\zpe)}\right)^{\frac{1}{3}} \nonumber
\end{equation}
and
\begin{equation}
\pmass(z;\mpe,\zpe)=\frac{\rhos(z)\, \rs^3(z)}{\rhos(\zpe)\, \rs^3(\zpe)}
\label{functionsPcPd}
\end{equation}
describe the evolution of the density and mass of the halo core between $z$ and $\zpe$ for a halo mass $\mpe$. Both quantities are normalized so that the limit of exact pseudo-evolution corresponds to unity.

In scale-free models, $\prho$ and $\pmass$ can only be a function of $\mpe/M_*$, or equivalently of $\nu/\nupe$. Thus, the equations can be written as
\begin{equation}
\frac{c}{ \left[g(c)\right]^{\frac{1}{3}}} = \frac{A}{\nu (\mpe,z)} \prho \left(\frac{\nu (\mpe,z)}{\nupe} \right) 
\label{relation-general-SF-1}
\end{equation}
and
\begin{equation}
\nu(M,z) = \nu(\mpe,z) \left[ \frac{g(c)}{g(\cpe)} \pmass \left( \frac{\nu (\mpe,z)}{\nupe} \right) \right]^{\frac{3+n}{6}} \,,
\label{relation-general-SF-2}
\end{equation}
where, as above, $A = \nupe F(\cpe)$ and $\prho(y)$ and $\pmass(y)$ are two dimensionless functions. We assume that they approach unity in the pseudo-evolving limit of $y \ll 1$. For any given functional form of $\prho(y)$ and $\pmass(y)$, we could numerically determine the \cm relation. We now insert Equation~(\ref{relation-general-SF-2}) into Equation~(\ref{relation-general-SF-1}) to obtain
\begin{equation}
\frac{c}{ \left[g(c)\right]^{(5+n)/6}}=  \frac{A}{\nu(M,z)} P^\prime \left(c, \frac{\nu (M,z)}{\nupe} \right) \,,
\label{relation-general-SF-3}
\end{equation}
where we have gathered the $\prho$ and $\pmass$ terms into a new function $P^\prime(c, y)$ which approaches unity for $y \ll 1$, and which we expect to monotonically increase. Exploring functional forms for $P^\prime(c, y)$, we find that the simulation results are well fit by a remarkably simple two-parameter extension of our analytic result for the pseudo-evolving limit,
\begin{equation}
\frac{c}{ \left[g(c)\right]^{(5+n)/6}} = \frac{A(n)}{\nu} \left( 1 + \frac{\nu^2}{B(n)} \right) \,,
\label{equ:fit_step3}
\end{equation}
where the constants $A(n)$ and $B(n)$ may depend on $n$ only. Once again, we parameterize $A$ and $B$ to linear order in $n$, 
$A = a_0 + a_1 (n + 3)$ and $B = b_0 + b_1 (n + 3)$, resulting in four free parameters. For $\ctoc$, we find the best-fit 
values $a_0 = 2.44$, $a_1 = 4.49$, $b_0 = 3.48$, and $b_1 = 7.46$ (we note that $a_0$ and $a_1$ take on different values from the fit in Section~\ref{sec:theory:pe_eds}). The right column of Figure~\ref{fig:fit_steps} demonstrates 
that Equation~(\ref{equ:fit_step3}) describes the data to 5\% except at the highest peak heights. When combining the 
scale-free and \LCDM data in the next section, we will see that, at very steep slopes $n$, the scale-free 
and \LCDM data are incompatible in that they behave differently at the same $\nu$ and $n$, regardless of 
how $\neff$ is defined. Thus, we ignore the issue at this point.

We have considered more general forms, for example, by including a variable exponent of $\nu$ in the correction term, but we find only marginally better fits. We note that we assumed that $P^\prime(c, y)$ was not a function of $c$, while one might
expect that the $g(c)$ term in Equation~(\ref{relation-general-SF-2}) would lead to such a dependence. For example, if we assume $\pmass(y) = 1$ (pure pseudo-evolution) and take a simple quadratic form for $\prho(y)$, then $B \propto g(c)^{(3+n)/3}$. We have experimented with such functions but find no improvement to the fits. The reason for the weak impact of this term is that, where the correction term dominates at high masses, $c$ and $g(c)$ vary relatively little. Thus, any such dependence is absorbed into the parameters $b_0$ and $b_1$.

\subsection{All Halos in \LCDM Cosmologies}
\label{sec:theory:all_lcdm}

\begin{deluxetable}{lccl}
\tablecaption{Best-fit parameters}
\tablehead{
\colhead{Par.} &
\colhead{Median } &
\colhead{Mean } &
\colhead{Explanation}
}
\startdata
\rule{0pt}{1ex} $\kappa$      & $0.41$  & $0.42$ & Loc. in $R$ where $\neff$ is computed \\
\rule{0pt}{0pt} $a_0$         & $2.45$  & $2.37$ & Normalization of $A$ \\
\rule{0pt}{0pt} $a_1$         & $1.82$  & $1.74$ & $\neff$-dependence of $A$ \\
\rule{0pt}{0pt} $b_0$         & $3.20$  & $3.39$ & Normalization of $B$ \\
\rule{0pt}{0pt} $b_1$         & $2.30$  & $1.82$ & $\neff$-dependence of $B$ \\ 
\rule{0pt}{0pt} $c_{\alpha}$  & $0.21$  & $0.20$ & $\alphaeff$-dependence
\enddata{}
\tablecomments{Best-fit parameters for the fitting function of Equation~(\ref{equ:fit_func}). The two sets of parameters refer to fits to the median and mean concentrations, respectively. The fits were performed using our full halo sample including all masses, redshifts, and cosmologies (see Section~\ref{sec:sims:sample})}.
\label{table:params}
\end{deluxetable}

\begin{figure*}
\centering
\includegraphics[trim =  0mm 3mm 2mm 0mm, clip, scale=0.69]{\figdir/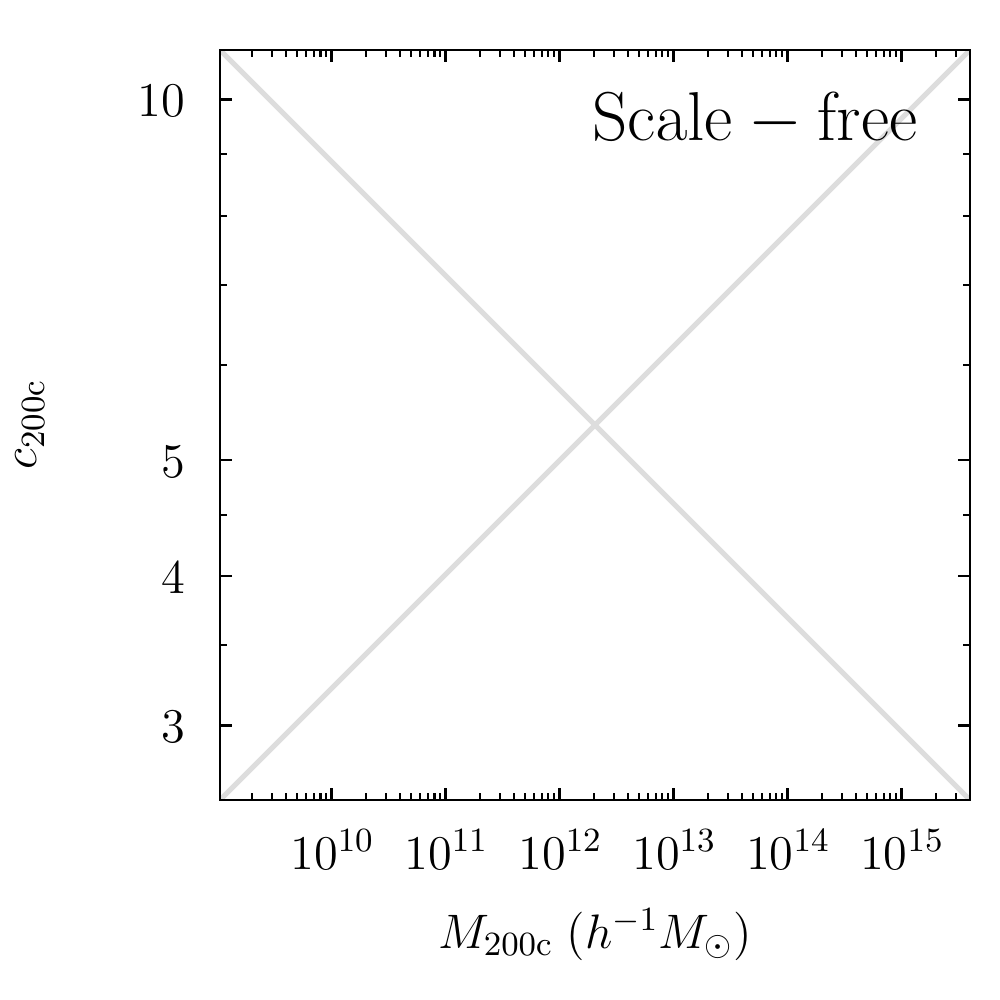}
\includegraphics[trim = 21mm 3mm 2mm 0mm, clip, scale=0.69]{\figdir/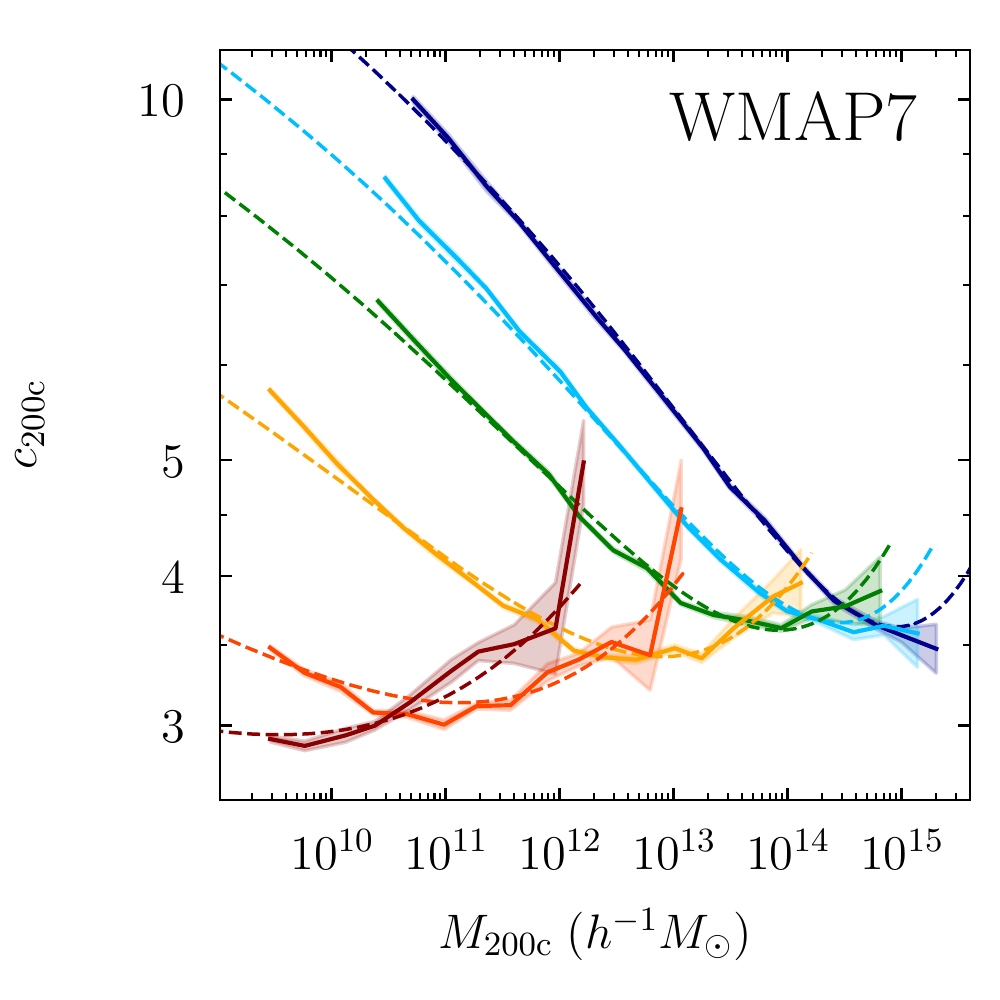}
\includegraphics[trim = 21mm 3mm 2mm 0mm, clip, scale=0.69]{\figdir/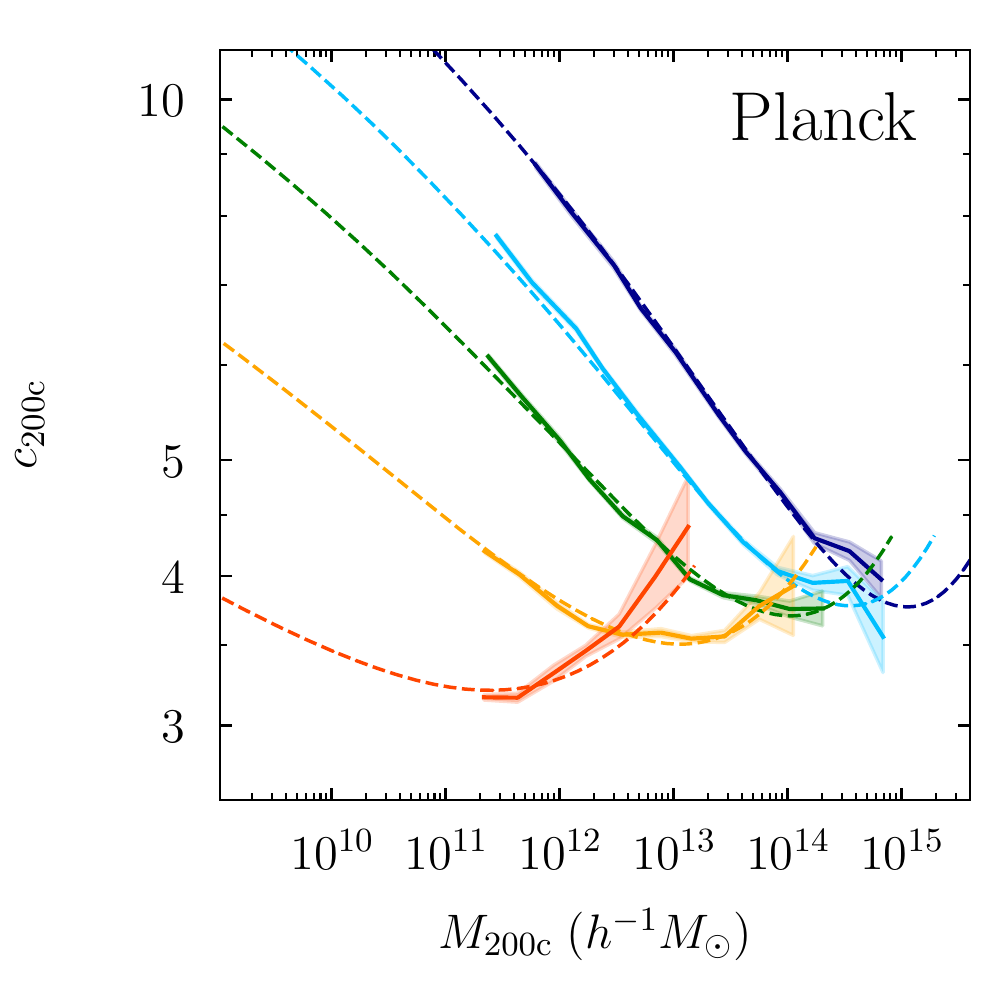}
\includegraphics[trim =  0mm 5mm 2mm 3mm, clip, scale=0.69]{\figdir/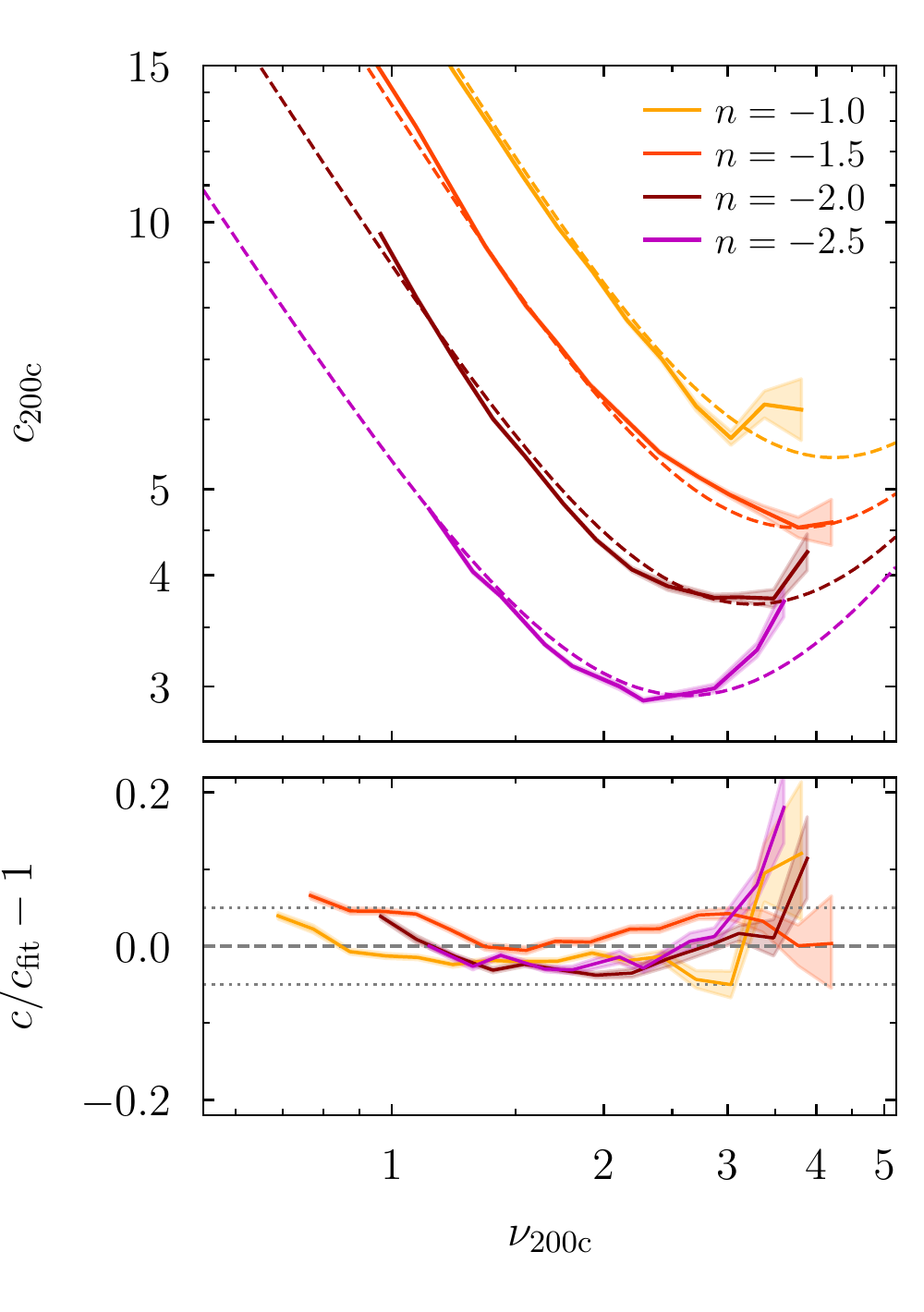}
\includegraphics[trim = 21mm 5mm 2mm 3mm, clip, scale=0.69]{\figdir/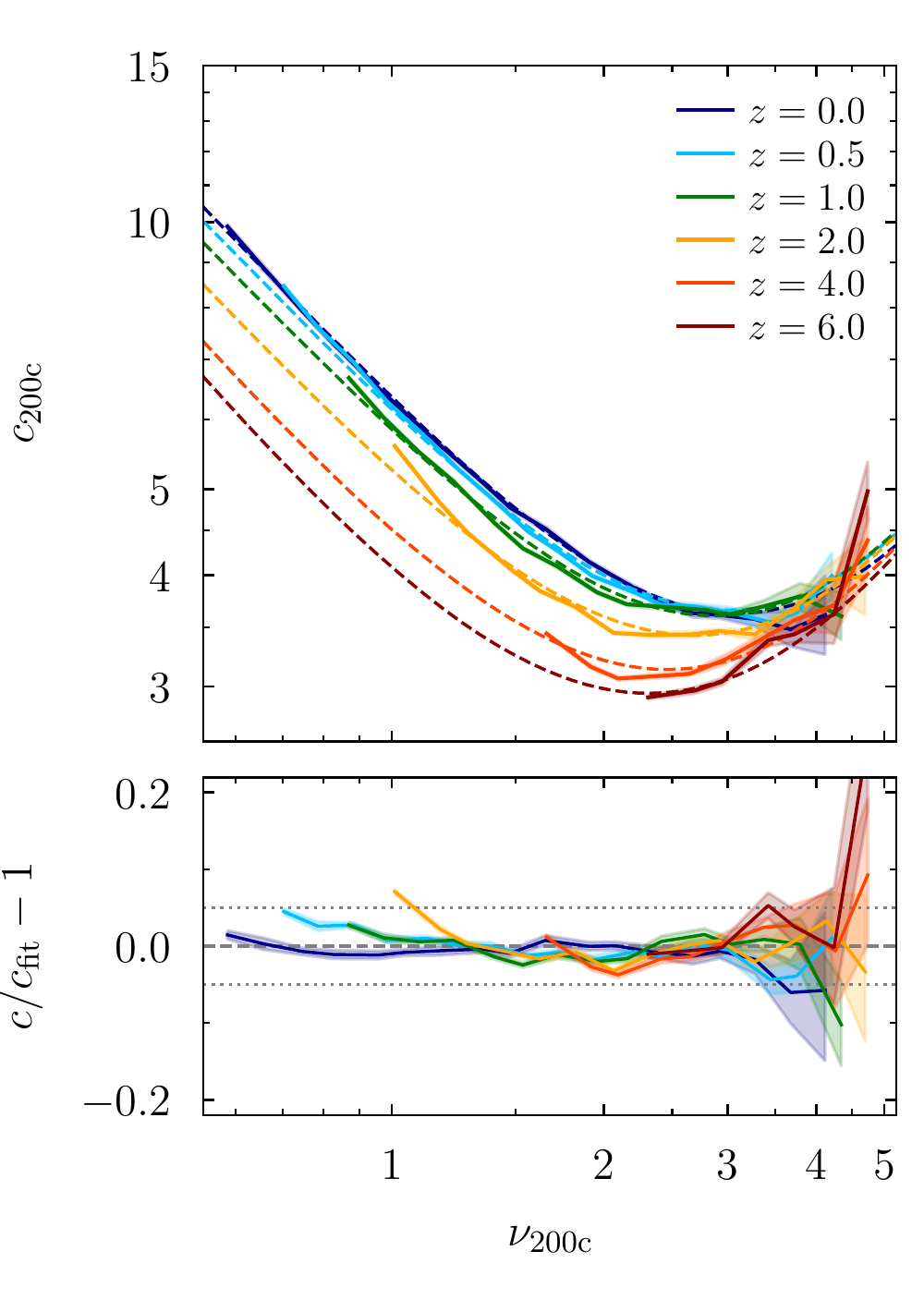}
\includegraphics[trim = 21mm 5mm 2mm 3mm, clip, scale=0.69]{\figdir/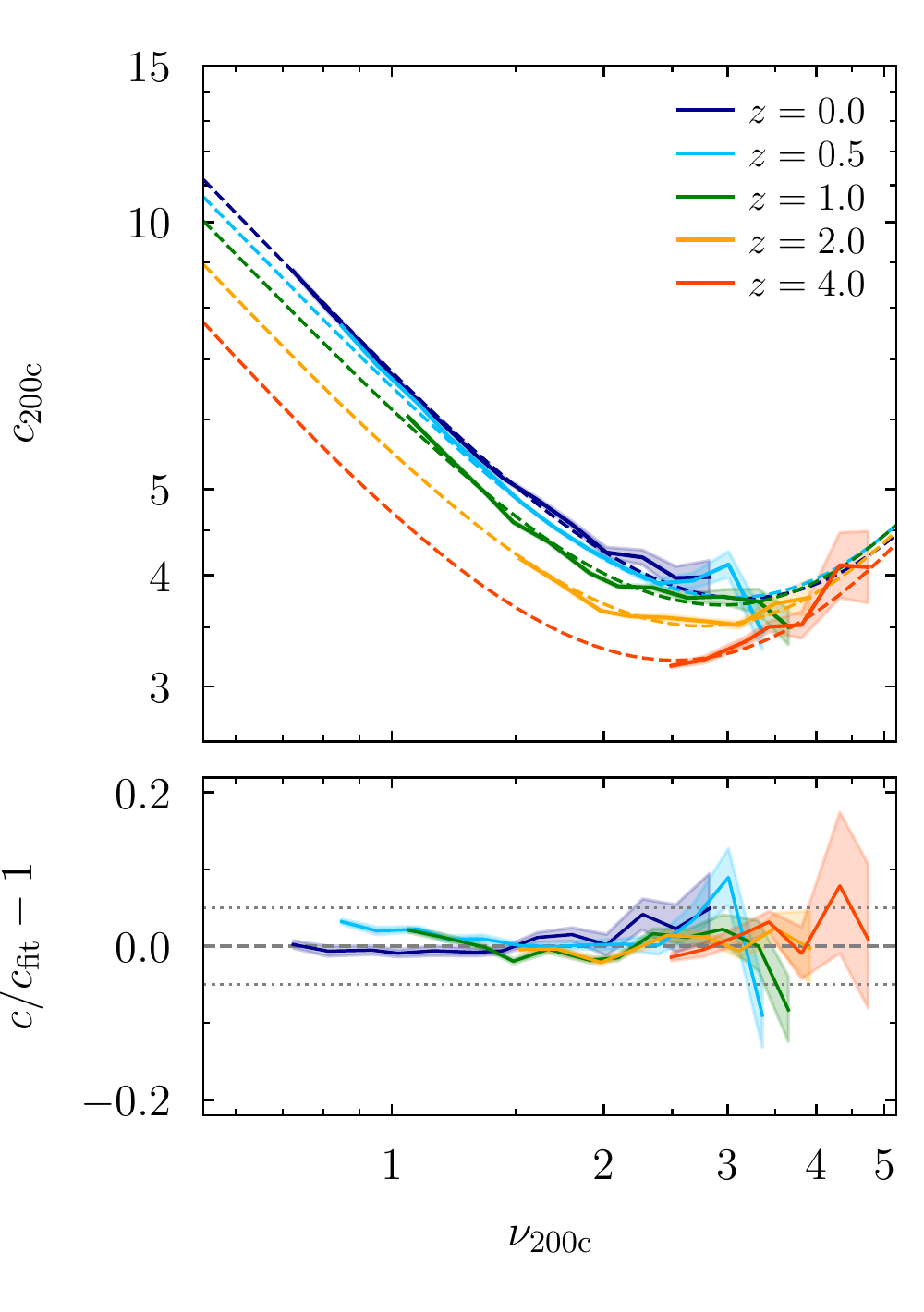}
\caption{Quality of our fitting function (Equation~\ref{equ:fit_func}). The top row shows the simulation data as solid lines in \cm space, the middle row shows the same data but in \cnu space, and the bottom row shows the relative difference between the data and fitting function in \cnu space. The columns refer to the scale-free simulations, the simulations with {\it WMAP7} cosmology, and the {\it Planck} simulations (Section~\ref{sec:sims}). For the scale-free models, halo mass is not a meaningful variable. The data shown in the middle row represent a compilation of data from different redshifts \citepalias{diemer_15}. The fitting functions match both the scale-free and \LCDM data to 5\% or better, with the exception of the high-$\nu$ end of the scale-free cosmologies.}
\label{fig:main_fit}
\end{figure*}

\begin{figure*}
\centering
\includegraphics[trim =  2mm 6mm 2mm 0mm, clip, scale=0.64]{\figdir/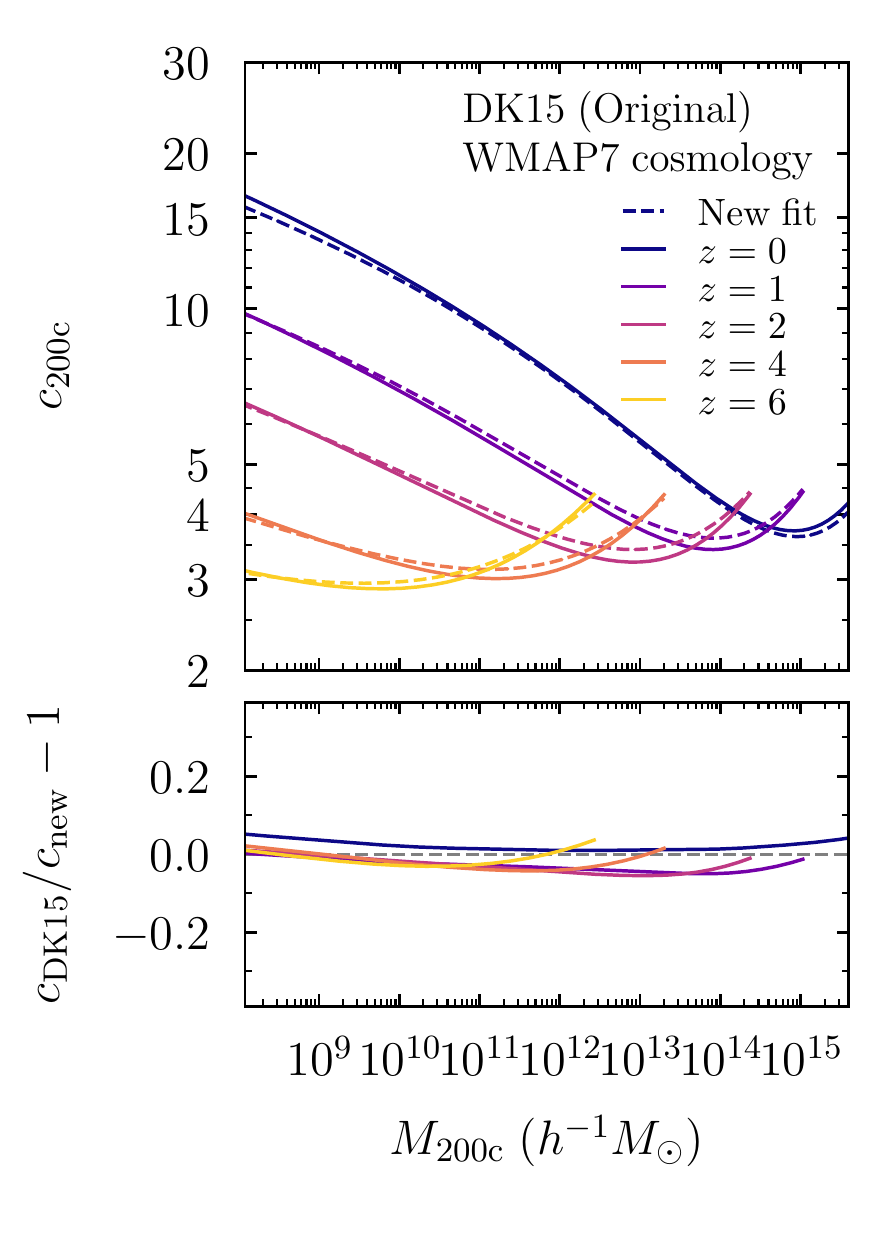}
\includegraphics[trim = 23mm 6mm 2mm 0mm, clip, scale=0.64]{\figdir/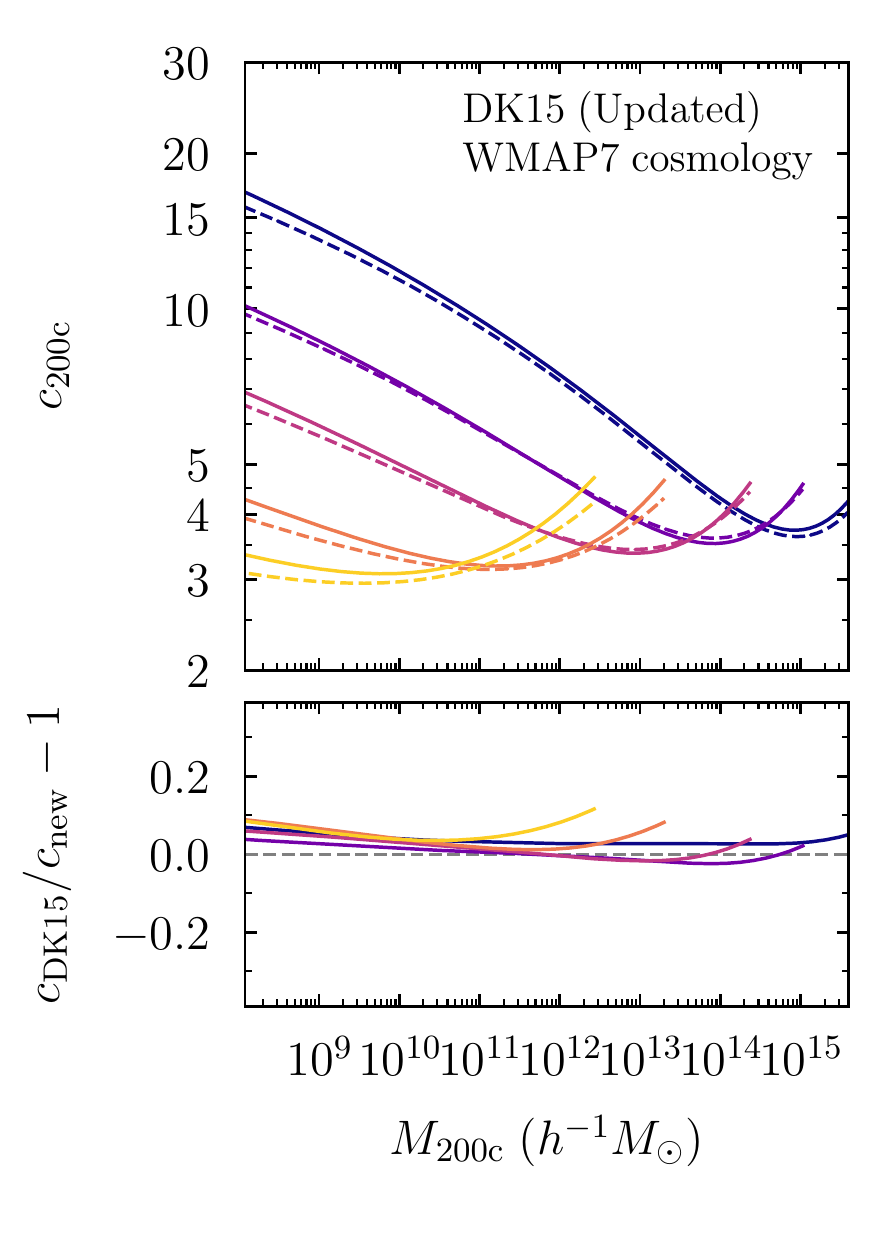}
\includegraphics[trim = 23mm 6mm 2mm 0mm, clip, scale=0.64]{\figdir/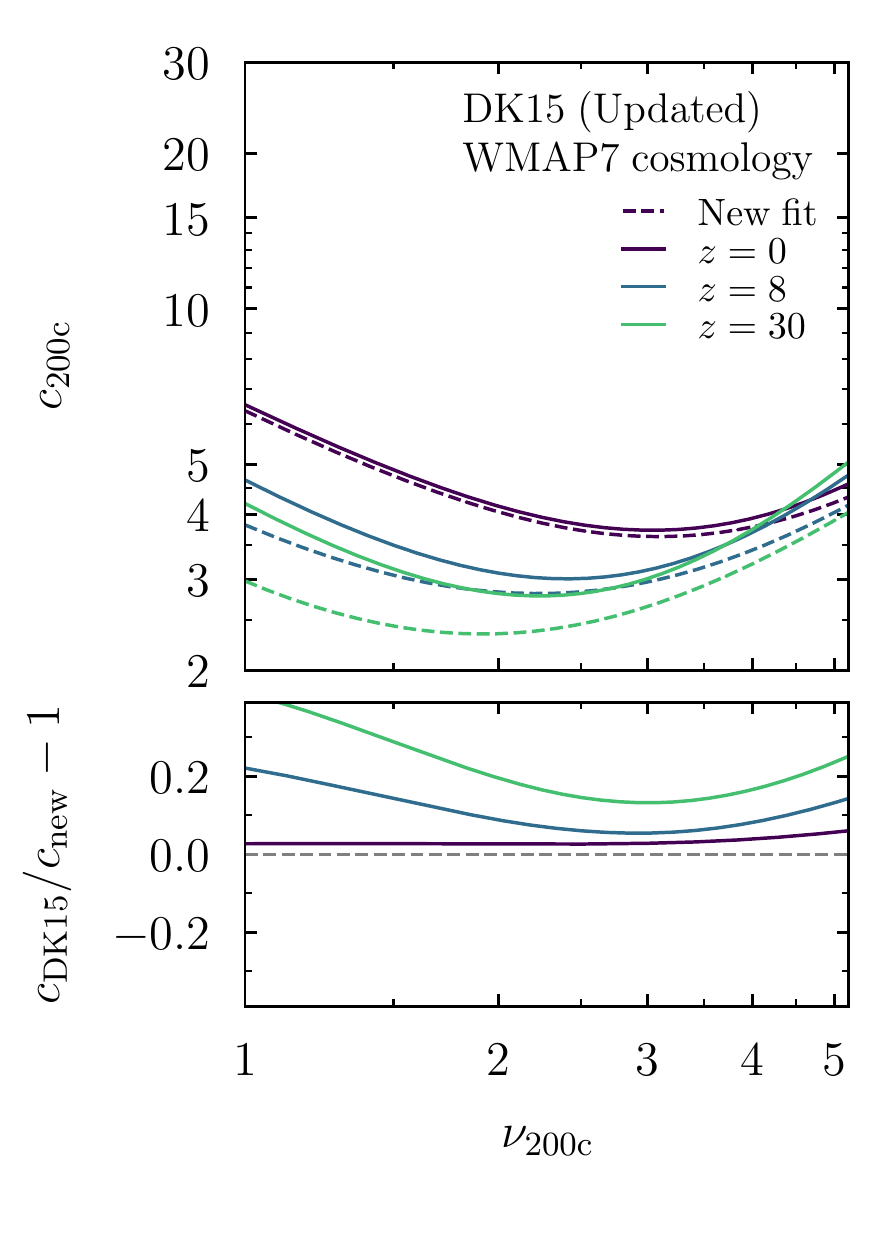}
\includegraphics[trim = 23mm 6mm 2mm 0mm, clip, scale=0.64]{\figdir/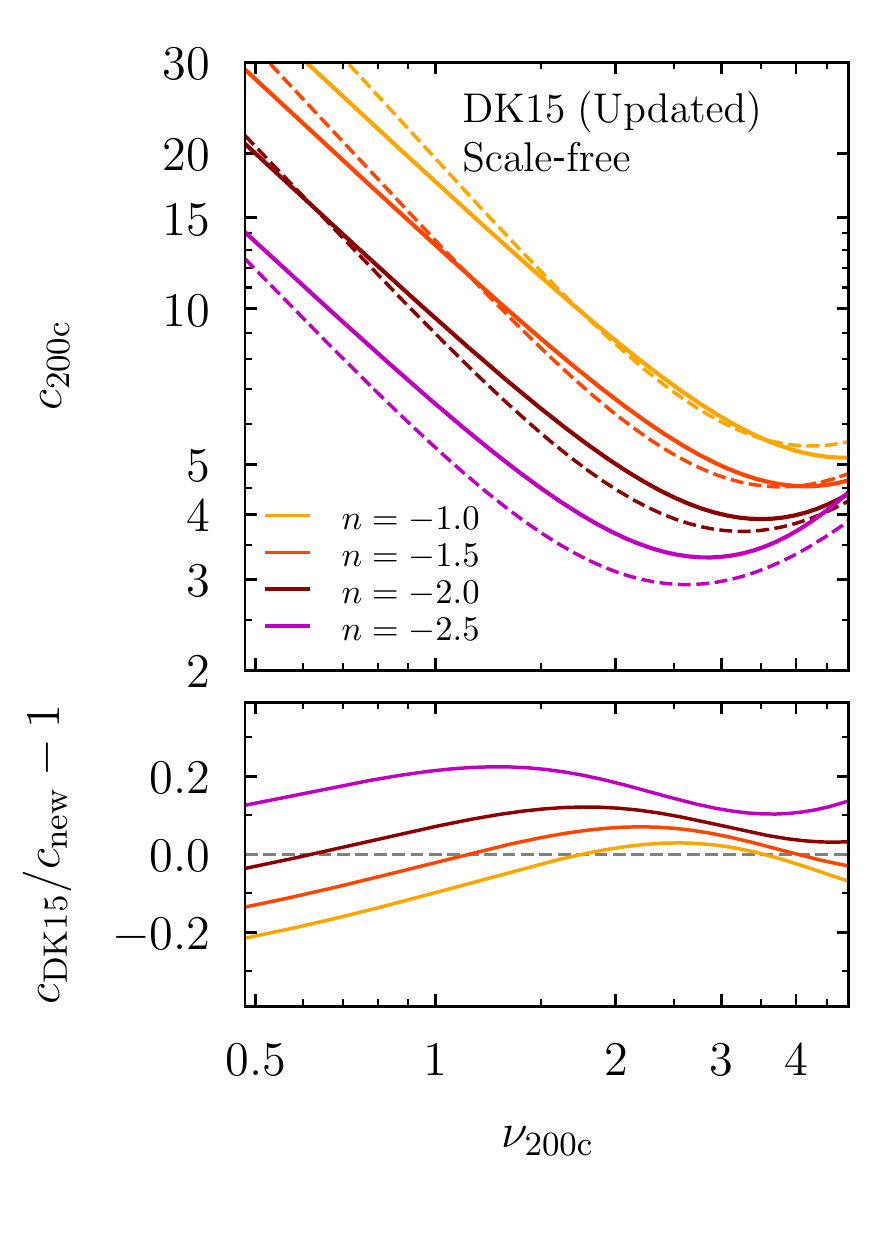}
\caption{Comparison of our model with \citetalias{diemer_15} in \LCDM (left three panels) and scale-free (right panel) cosmologies. In each panel, the dashed lines show our new model, the solid lines \citetalias{diemer_15}. {\it Left two panels:} the \cm relations for both the original and updated \citetalias{diemer_15} parameterizations (Appendix~\ref{sec:app:dk15update}) exhibit modest differences, about 5\%. {\it Third panel:} at very high redshift, the new model predicts concentrations up to 40\% lower. For reference, at this redshift an Earth mass corresponds to $\nu \approx 2$. {\it Right panel:} the differences with the \citetalias{diemer_15} model are much larger for scale-free models than for \LCDM, about 20\% over a wide range of peak heights.}
\label{fig:model_comp_dk15}
\end{figure*}

We now attempt to generalize ansatz~(\ref{equ:fit_step3}) to \LCDM cosmologies. A first approach might be to return to the general expression~(\ref{equ:c_pe_convenient}), extend it with simple forms of $\prho$ and $\pmass$ that depend on $(1+z)/(1+\zpe)$, and to enforce that the expressions reduce to our fitting function from the last section in the scale-free limit. By construction, we recover the pseudo-evolving limit if $\zpe \gg z$ (where $\zpe$ is defined as in Equation~\ref{equ:zpe}). However, for masses greater than $\mpe$, we encounter a serious problem: the redshift at which the halo will start to pseudo-evolve lies in the future. In a scale-free cosmology, this is unproblematic because we can use self-similarity, i.e., $(1+z)/(1+\zpe) = \nu(\mpe,z) / \nupe$ where $\nupe$ is a constant (Section~\ref{sec:theory:pe_eds}). In \LCDM, assuming that $\nupe$ is constant means that some large halos will never pseudo-evolve because the linear growth factor $D(z)$ asymptotes to a finite value that large-$\nu$ halos will never reach. This conclusion is manifestly incorrect; in reality, we expect the opposite to occur. The rapid expansion of the universe will stop all physical mass accretion and eventually lead to pure pseudo-evolution. Thus, the assumption that $\nupe$ is a constant can only be valid in cosmologies with expansion histories similar to EdS, leading us to abandon this approach.

Instead, we return to the question that first led \citetalias{diemer_15} to introduce a dependence of concentration on $n$: what breaks the self-similarity of the \cnu relation? There are three factors: the scale-dependent power spectrum, the halo selection criterion (i.e., the spherical overdensity), and the non-EdS expansion history of the universe. The first effect, we have argued, should be taken into account via the dependence on $\neff$. The second effect is taken into account in our pseudo-evolution calculation and should be subdominant for rapidly accreting halos. The third effect, however, could influence halo concentrations, though we have no a priori insight into its quantitative impact. Further motivation to explore the physics of non-EdS expansion is provided by our fitting results: when fitting all data (including \LCDM) with Equation~(\ref{equ:fit_step3}), we observe that the low-redshift \LCDM data prefer a lower \cmr compared to the scale-free and high-$z$ data. Thus, we wish to construct a physical variable that captures the impact of the expansion history, similar to the way $\neff$ captures the effect of the shape of the power spectrum. For this purpose, we define the {\it effective exponent of linear growth},
\begin{equation}
\alphaeff (z) = -\frac{d \ln D(z)}{d \ln (1+z)} \,.
\label{equ:alphaeff}
\end{equation}
This function is a natural choice for characterizing the effect of deviations from a standard EdS cosmology on the evolution of structure \citep[used, for example, in the context of modified theories of gravity; e.g.][]{carroll_06, linder_18}. As discussed in Section~\ref{sec:discussion}, our choice also has an additional motivation in the context of scale-free cosmological models. The bottom panel of Figure~\ref{fig:neff_alphaeff} shows $\alphaeff$ as a function of redshift. At $z \gsim 2$, the expansion is EdS-like and $\alphaeff = 1$. At low redshifts, the growth factor evolves more slowly, reaching a slope of $\alphaeff \approx 0.5$ at $z = 0$. In the future, $\alphaeff$ will approach zero.

Based on $\alphaeff$, we propose one of the simplest possible extensions of the fitting function from the previous section: we assert that all concentrations are modified by a factor that linearly depends on $\alphaeff$. For convenience, we rewrite Equation~(\ref{equ:fit_step3}) in terms of $ \tilde{G}(x)$, the inverse of
\begin{equation}
G(x) = \frac{x}{ \left[g(x)\right]^{(5+\neff)/6}} \,,
\end{equation}
and multiply it by an $\alphaeff$-dependent term $C(\alphaeff)$, 
\begin{equation}
\label{equ:fit_func}
c = C(\alphaeff) \times \tilde{G} \left( \frac{A(n_{\rm eff})}{\nu} 
\left[ 1+\frac{\nu^{2}}{B(n_{\rm eff})} \right]\right) \,,
\end{equation}
where
\begin{eqnarray}
\label{equ:fit_func2}
A(\neff) &=& a_0 (1 + a_1 (\neff + 3)) \, \nonumber \\
B(\neff) &=& b_0 (1 + b_1 (\neff + 3)) \, \nonumber \\
C(\alphaeff) &=& 1 - c_\alpha (1 - \alphaeff) \,.
\end{eqnarray} 
Including $\kappa$, our function now has six free parameters, whose best-fit values are listed in Table~\ref{table:params}. Figure~\ref{fig:main_fit} demonstrates the quality of Equation~(\ref{equ:fit_func}) in a simultaneous fit to the $\ctoc$ data from all our simulations. The function fits the data to 5\% or better at virtually all redshifts and power spectrum slopes (taking into account the statistical uncertainties on bins with few halos). Figure~\ref{fig:main_fit} shows fits to the median \cmr, but the mean relations are fit to similar accuracy. Fits to other mass definitions, e.g. $\ctom$, result in poorer fits than $\ctoc$. Instead, we numerically convert the $\ctoc$ results to other definitions a posteriori (Appendix~\ref{sec:app:mdef}).

Unlike \citetalias{diemer_15}, we assign the same weight to all cosmologies and redshifts. This simplification is possible because there is little tension between the \LCDM and scale-free data except at high peak heights and steep slopes, $n \approx -2.5$, where the scale-free and \LCDM data are slightly incompatible. In this regime, $\alphaeff = 1$ in both cosmologies, meaning that a dependence on the expansion history cannot resolve the disagreement.

Remarkably, our fitting function improves on the quality of the \citetalias{diemer_15} fit while using one fewer free parameter (six rather than seven). The improvement is most notable in the scale-free models, where the updated \citetalias{diemer_15} model disagrees with the data by up to 20\% over a wide range of $n$ and $\nu$ (Figure~\ref{fig:dk15update}). Most importantly, this improvement was achieved by predicting the low-mass shape of the \cnur from first principles and by adding a physically meaningful variable, $\alphaeff$.


\section{Comparison with Other Models and Data}
\label{sec:comparison}

We have derived a semianalytical fitting function for the \cmr and demonstrated that it fits our simulation data accurately. 
In this section, we compare our new model to previously published fitting functions and simulation data.

\subsection{Comparison with DK15}
\label{sec:comparison:dk15}

Figure~\ref{fig:model_comp_dk15} shows a detailed comparison of our model to that of \citetalias{diemer_15} for both \LCDM and scale-free cosmologies and from $z = 0$ to $z = 30$. In Appendix~\ref{sec:app:dk15update}, we present slightly adjusted parameters for the \citetalias{diemer_15} model that correspond to a shift in the data due to a small numerical error in the original paper. The left two panels of Figure~\ref{fig:model_comp_dk15} compare the new model to both those versions of \citetalias{diemer_15} for the {\it WMAP7} cosmology and at the redshifts where the models were constrained by \LCDM simulation data, $0 < z < 6$. Both versions of the \citetalias{diemer_15} model agree with the new function to 5\% or better, an agreement that is expected given that both models fit the same data. 

A good agreement at low redshift, however, does not guarantee that the models extrapolate to extremely high redshifts in the same fashion \citep[e.g.,][]{ludlow_14}. Thus, we compare the models up to $z = 30$ in the third panel of Figure~\ref{fig:model_comp_dk15}. Here we show the \cnu rather than the \cmr because the mass range considered varies strongly with redshift. For example, $\nu = 1$ corresponds to $10^{-17}$ solar masses at $z = 30$. At high redshifts, we notice much larger disagreements between the models, up to 40\%, with the new model predicting consistently lower concentrations. This difference is related to the new model's good fit to the $n = -2.5$ scale-free cosmology, as the power spectrum slope gets very steep at high redshift. Because of this connection, the high-$z$ predictions of our model are not entirely unconstrained, even though they were not directly trained on \LCDM data. Our model naturally reproduces the very small concentrations that are found in detailed simulations at high redshift \citep[$c \approx 2$--$3$ for halos with $\nu\approx 2$ at $z = 30$,][compare to Figure 8 in \citetalias{diemer_15}]{diemand_05, anderhalden_13, ishiyama_14}.

As expected, the right panel of Figure~\ref{fig:model_comp_dk15} demonstrates that the new model and \citetalias{diemer_15} differ by up to 20\% in their predictions for the concentrations in scale-free cosmologies, with the new model producing a much better fit. As discussed in Section~\ref{sec:theory}, this is largely a product of the careful consideration of the physics that distinguish EdS and non-EdS universes, namely, the definitions of $\neff$ and $\alphaeff$ and their effect on the \cnur.

\subsection{Comparison with Previous Models}
\label{sec:comparison:other}

\begin{figure*}
\centering
\includegraphics[trim =  2mm  6mm 2mm 0mm, clip, scale=0.64]{\figdir/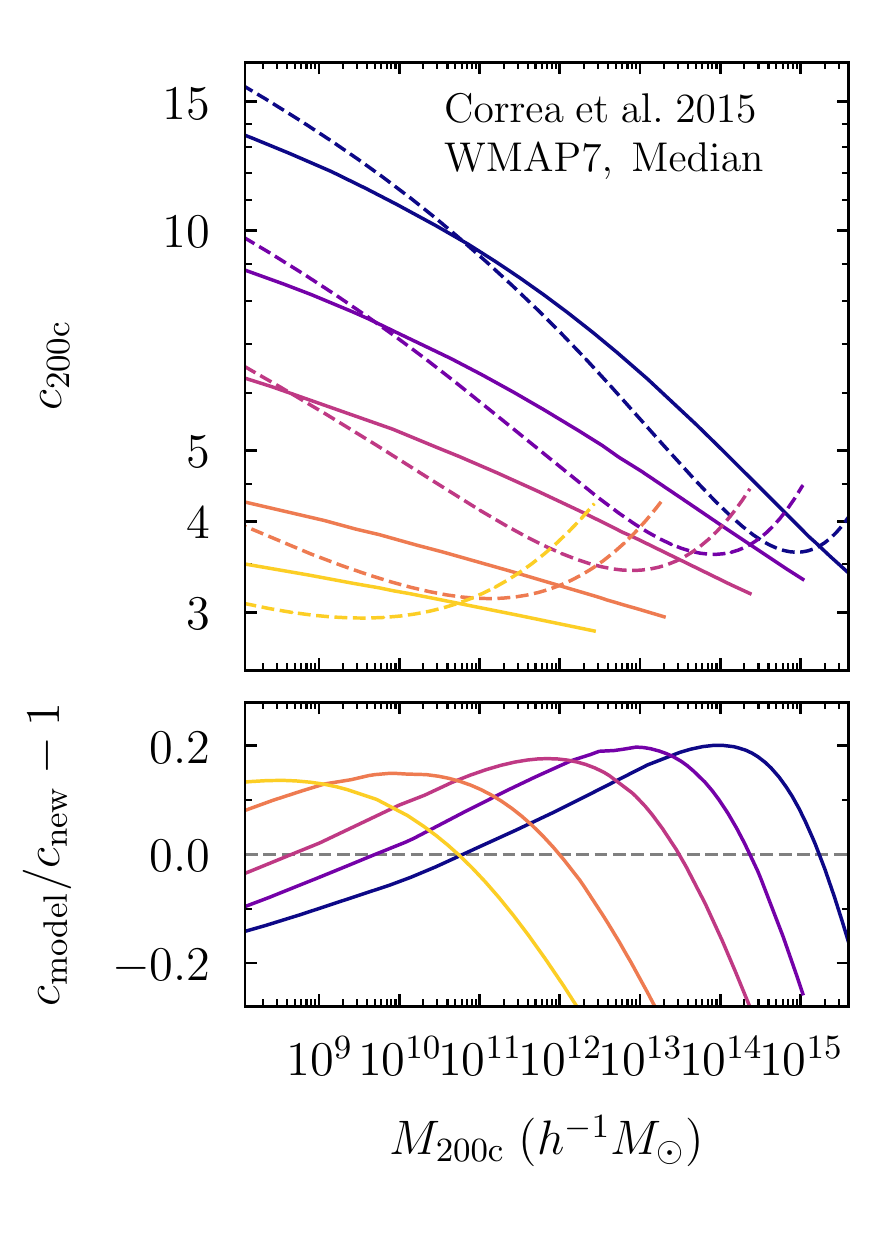}
\includegraphics[trim = 23mm  6mm 2mm 0mm, clip, scale=0.64]{\figdir/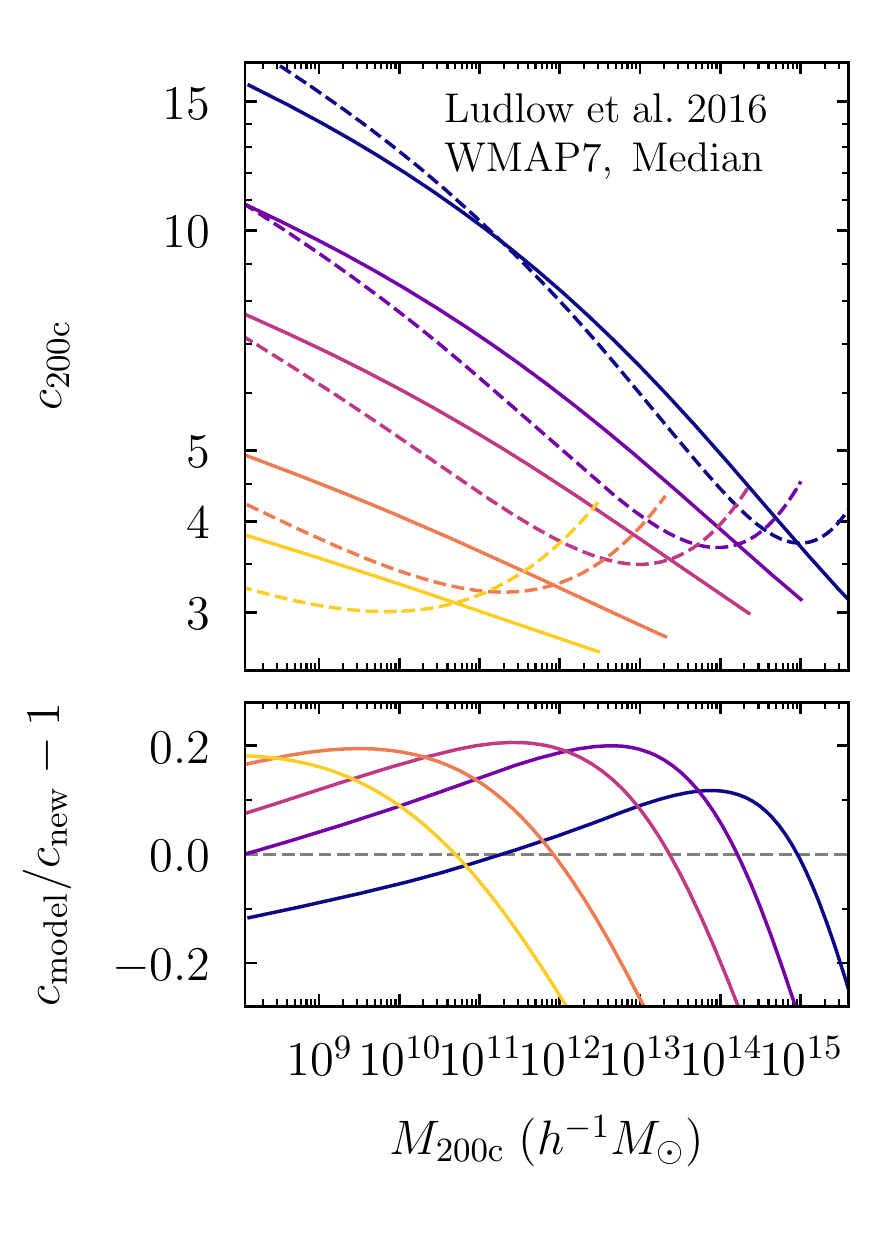}
\includegraphics[trim = 23mm  6mm 2mm 0mm, clip, scale=0.64]{\figdir/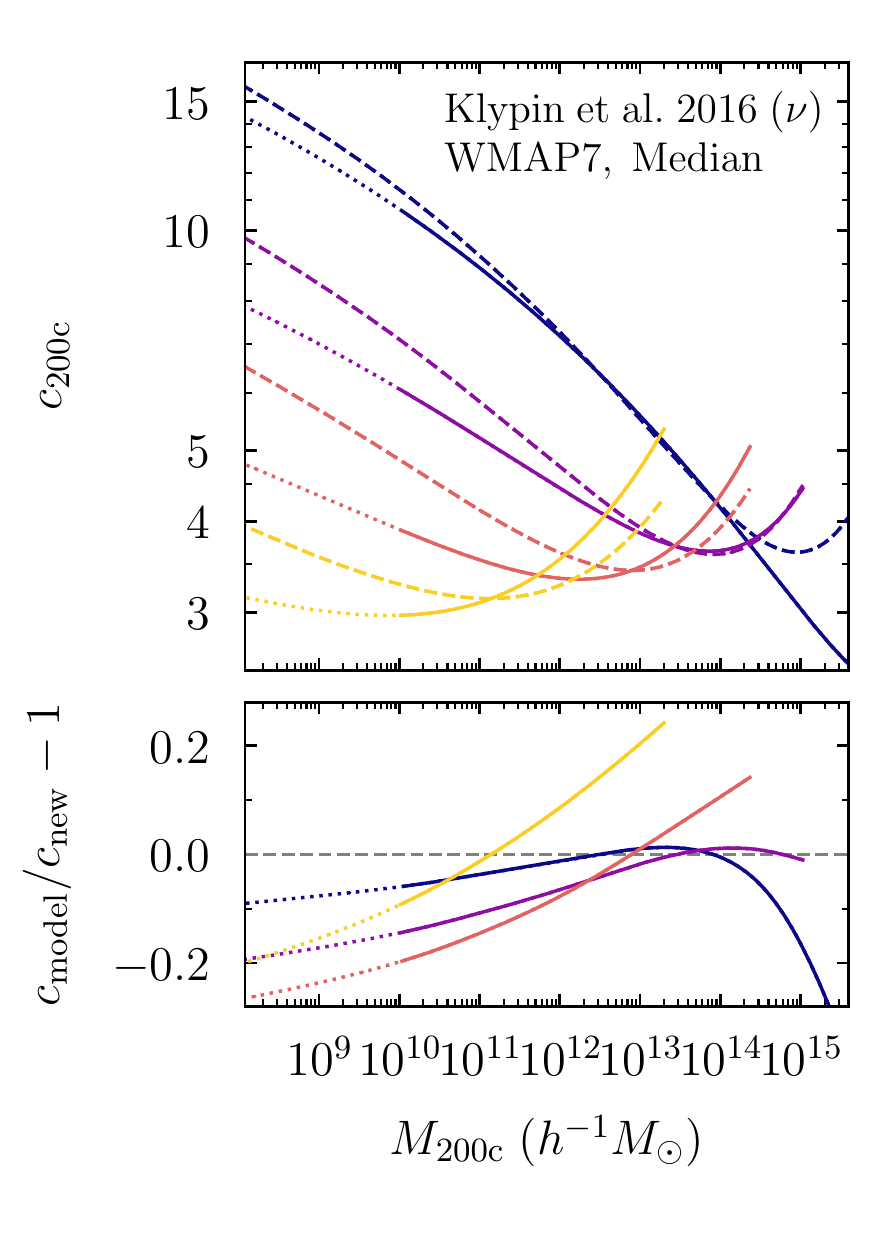}
\includegraphics[trim = 23mm  6mm 2mm 0mm, clip, scale=0.64]{\figdir/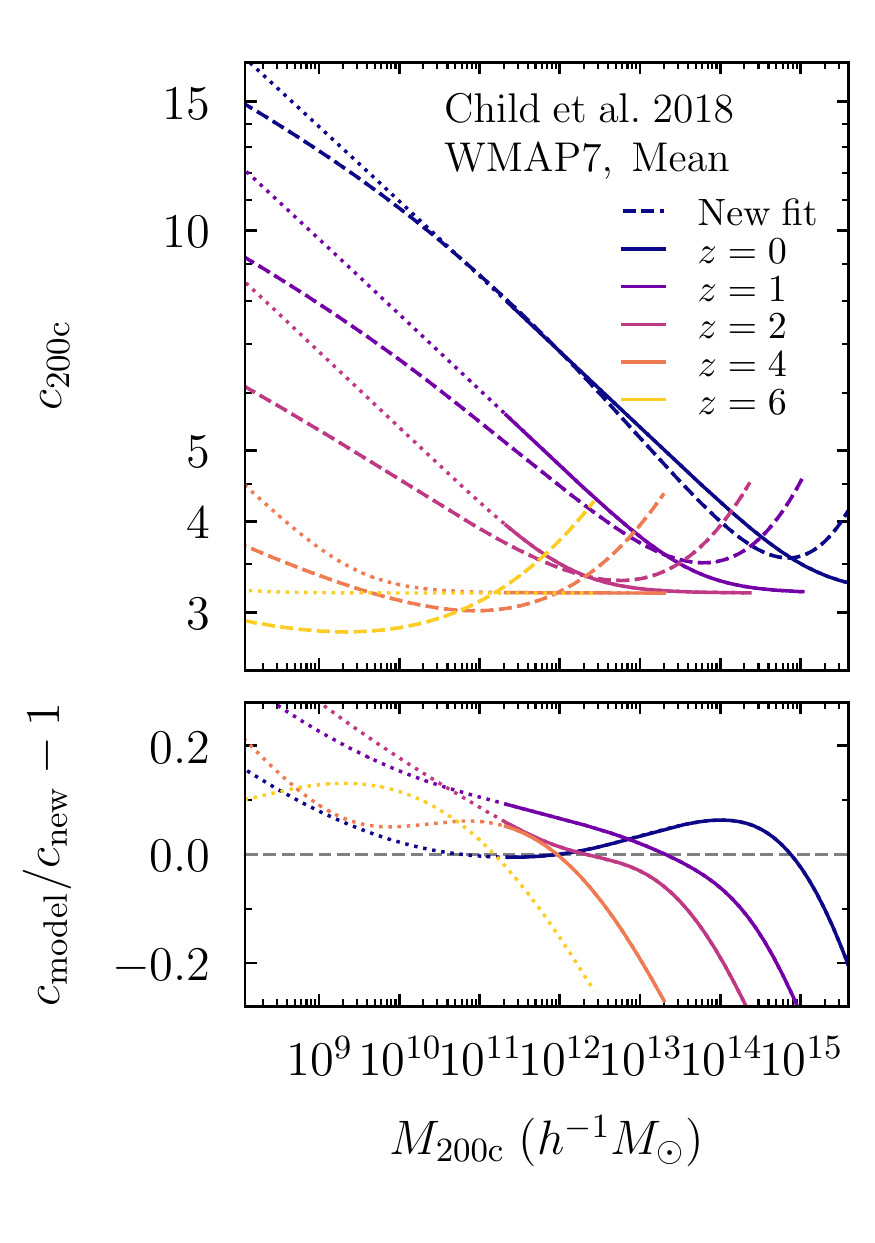}
\caption{Comparison between our new model and previous works, evaluated for the {\it WMAP7} cosmology used in our simulations. In each set of panels, the top panel shows our function as dashed lines and the other model as solid lines, and the bottom panel shows the relative difference. We show fits to the mean or median \cmr, depending on the data used in the respective works. For the \citet{klypin_16} and \citet{child_18} fitting functions, the solid lines show the range where the fits were constrained, and the dotted lines show extrapolations of the fits beyond those regions. See Section~\ref{sec:comparison:other} for a detailed discussion.}
\label{fig:model_comp_other}
\end{figure*}

\citetalias{diemer_15} compared their model to those of \citet{bullock_01}, \citet{eke_01}, \citet{zhao_09}, \citet{prada_12}, \citet{bhattacharya_13}, \citet{ludlow_14}, and \citet{dutton_14}. Given the good agreement between our new function and  \citetalias{diemer_15} at $z \lsim 6$, we do not repeat those comparisons and refer the reader to Figures 9 and 10 in \citetalias{diemer_15}. Instead, we focus on models that have emerged since. Figure~\ref{fig:model_comp_other} shows a comparison of our model (dashed lines) with the models of \citet{correa_15_c} and \citet{ludlow_16}, as well as the fitting functions of \citet{klypin_16} and \citet{child_18}. Our model systematically differs from all others.

The first two models shown in Figure~\ref{fig:model_comp_other}, \citet{correa_15_c} and \citet{ludlow_16}, are based on the idea that the density profile of halos can be interpreted as their accretion history in units of the critical density of the universe at the time when certain shells of dark matter were accreted \citep{ludlow_13}. Based on their model for mass accretion histories \citep{correa_15_a, correa_15_b}, \citet{correa_15_c} compute the predicted median concentration. In such models, younger halos always have lower concentration, meaning that there is no upturn at high masses and that the model describes only relaxed halos. Their model differs from our function by up to 20\% in the mass and redshift range shown in Figure~\ref{fig:model_comp_other}.

In their original work, \citet{ludlow_13} directly connected the concentration of the mass accretion history (measured by fitting an NFW profile to the main progenitor branch) to the concentration of the density profile using a fitting function, an approach that was refined in \citet{ludlow_14}. \citet{ludlow_16} reverted to a simpler modeling of concentration using an approach based on the formation time similar to that of \citet[][see also \citealt{eke_01}]{navarro_97}. Here, the average collapse redshift of halos is computed based on Press--Schechter theory \citep{press_74, bond_91} and concentration is derived from the age of the halo. Given that the \citet{correa_15_c} and \citet{ludlow_16} functions are based on the same underlying logic, it is not surprising that they agree well and thus exhibit similar residuals to our model (see also Figure A1 in \citealt{ludlow_16}).

\begin{figure*}
\centering
\includegraphics[trim =  0mm 3mm 2mm 0mm, clip, scale=0.69]{\figdir/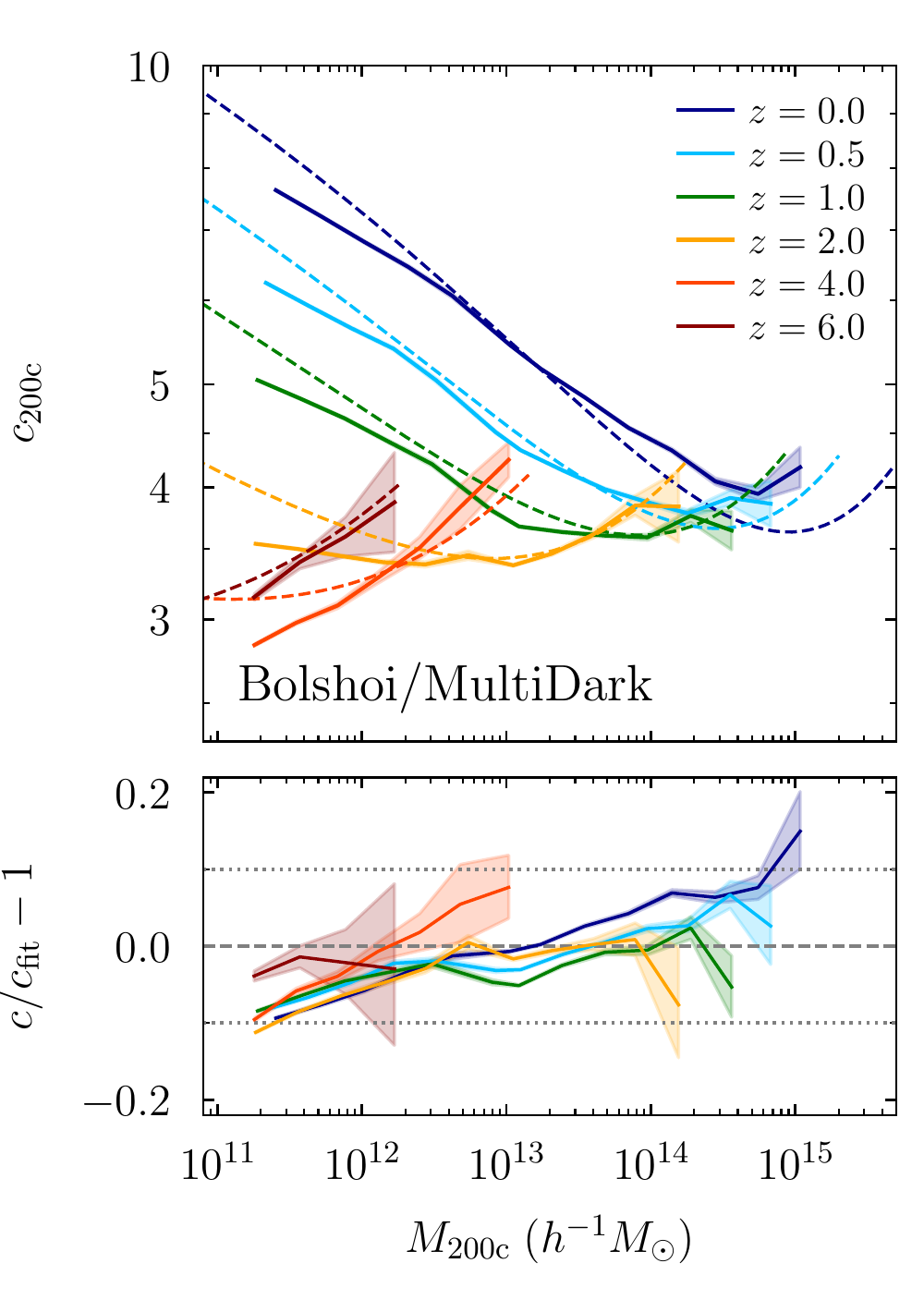}
\includegraphics[trim = 20mm 3mm 2mm 0mm, clip, scale=0.69]{\figdir/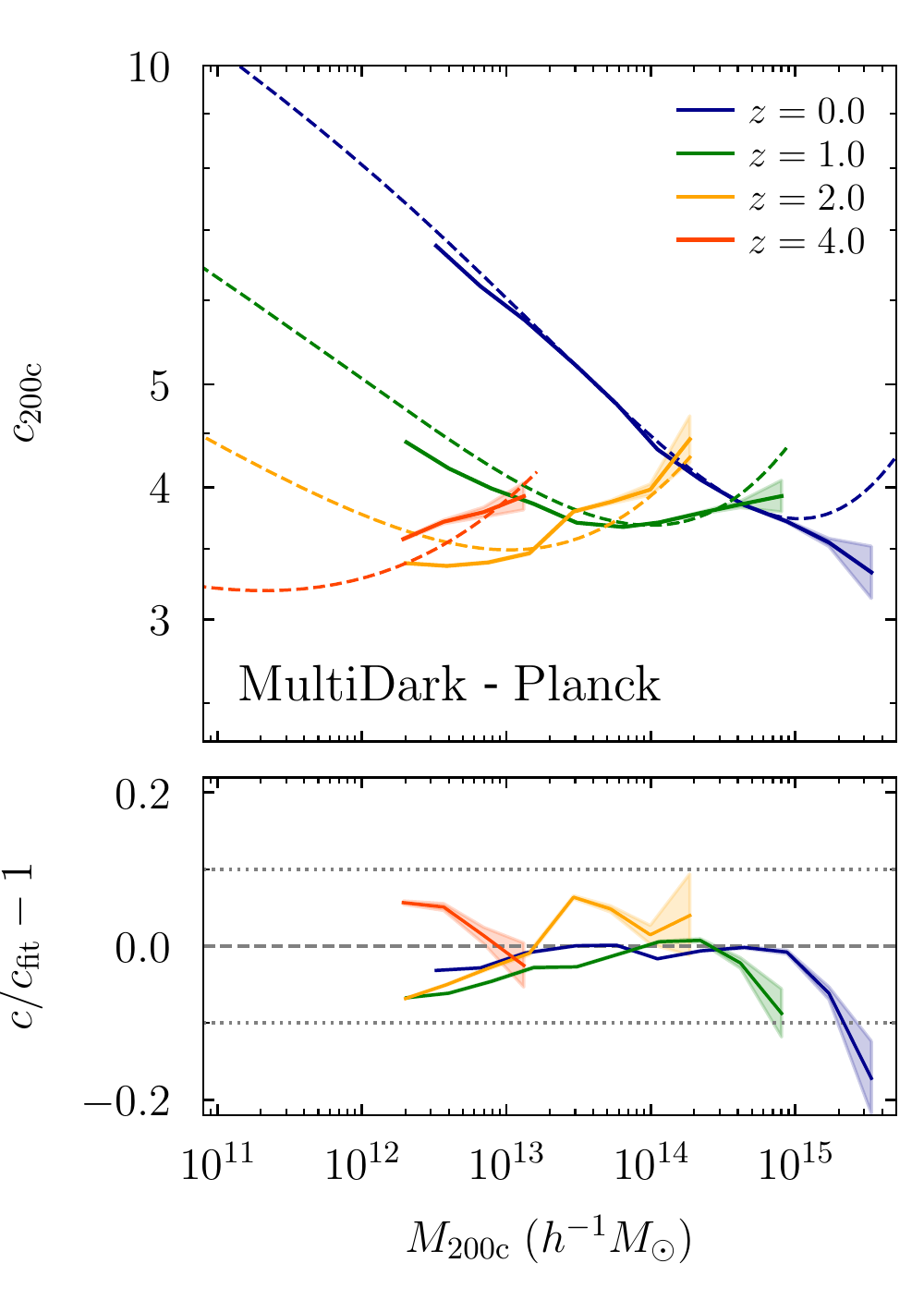}
\caption{Comparison of our fitting function (dashed lines) to data from other simulation suites. The left panel shows the median \cmr from the {\it Bolshoi} and {\it MultiDark} simulations \citep[which use the same {\it WMAP7} cosmology as some of our simulations;][]{klypin_11, prada_12}, the right panel shows results from the {\it MultiDark-Planck} simulation suite (\citealt{klypin_16}). All concentrations were computed by the \textsc{Rockstar} halo finder. Compared to {\it Bolshoi/MultiDark}, we find that our simulated concentrations follow a steeper \cmr at all redshifts, leading to about 10\% differences at low and high masses. Surprisingly, this trend does not appear to be present when comparing to the {\it MultiDark-Planck} suite, which is fitted to better than 10\% at all masses except for very large halos at $z = 0$, where the disagreement reaches about 15\%. Here, the data show no upturn, in agreement with our {\it Planck} data (Figure~\ref{fig:main_fit}) but in slight disagreement with {\it Bolshoi/MultiDark}. See Section~\ref{sec:comparison:data} for a detailed discussion.}
\label{fig:comp_multidark}
\end{figure*}

At the low-mass end, models that are directly based on the mass accretion history of halos such as \citet{correa_15_c} implicitly evolve concentration according to pseudo-evolution. The slow evolution of halos at late times (i.e., long after their formation redshift) is well matched with the expectation of a density profile that is static in physical units \citep{diemer_13_pe, more_15}. In the language of \citet{ludlow_13}, the steep outer NFW profiles of halos are caused by the shallow inner part of the NFW profile that describes their accretion histories. Our model makes this evolution analytically explicit at the low-mass end.

The last two models shown in Figure~\ref{fig:model_comp_other} are recently published empirical fitting functions to large numerical datasets. Based on the {\it MultiDark} simulations, \citet{klypin_16} proposed fitting functions with respect to both mass and peak height, as well as for relaxed and all halos (we choose the latter to match our halo selection). Both their fitting functions include a very strong upturn at high redshift and no upturn at $z = 0$, leading to 20\% disagreements with our model \citep[see also][]{klypin_11, prada_12, ludlow_12, meneghetti_13}. \citet{klypin_16} used the Bound Density Maxima halo finder \citep[BDM;][]{klypin_97} rather than \textsc{Rockstar} as well as a different algorithm to determine concentrations, which may explain an overall offset in the normalization of the concentrations \citep[e.g.,][]{dooley_14}. However, the differences also show a strong dependence on halo mass and redshift, indicating that other systematic effects are at play (as discussed in detail in Section~\ref{sec:comparison:data}).

Based on large $N$-body simulations \citep{heitmann_15, habib_16}, \citet{child_18} give numerous fitting functions based on different functional forms and halo samples. We compare our model to their formulation as a function of $M/M_*$ for all individual halos (as opposed to stacked or relaxed halos). Like the model of \citet{zhao_09}, their model presupposes that concentration reaches a floor of $\ctoc \approx 3$ at the highest masses, a prediction not borne out by our simulation data and thus in conflict with our model. 

While the \citet{klypin_16} and \citet{child_18} fitting functions are shown only in the mass range where they were constrained, we emphasize that they extrapolate to low masses differently: whereas peak-height-based fits will extrapolate similarly to our model, mass-based fits reach arbitrarily high concentrations \citep[e.g.,][]{ludlow_14}. This holds true even if the variable considered is $M/M_*$ as in \citet{child_18}.

\subsection{Comparison with Other Simulations}
\label{sec:comparison:data}

In general, the differences between \cm models arise not because of poor fits but because of differences in the underlying data. Such disagreements could be caused by the halo selection, different halo finders, the way concentration is measured from the density profile, and numerical effects that influence the density profiles themselves.

The first possible cause of differences, sample selection, was already discussed in Section~\ref{sec:comparison:other}: excluding unrelaxed halos leads to higher concentrations and no upturn at high peak height. This effect can partially explain the disagreements with the \citet{correa_15_c} and \citet{ludlow_16} models, but not those with the \citet{klypin_16} model. Another potential culprit is the method used to measure concentration. Like in most works in the literature, our concentrations were derived by fitting the NFW formula to halo density profiles, but there are alternatives. For example, computing $c$ based on circular velocities leads to a somewhat altered \cmr \citep{klypin_11, prada_12, meneghetti_13}. Even the details of the fitting procedure matter, for example, whether the density or mass profiles are fit \citep{povedaruiz_16} and out to what radius the density profile is considered. In our case, \textsc{Rockstar} fits the profile to $\rvir$. When larger radii are used, some particles can be missed because they are not part of the friends-of-friends group, biasing the measured scale radius high (\citealt{more_11_fof}; \citealt{behroozi_13_rockstar}; P. Mansfield \& A. Kravtsov 2019, in preparation).

To assess whether differences in the concentration measurement or in the underlying simulations are to be blamed for the disagreement with the \citet{klypin_16} model, we consider concentrations extracted from the {\it MultiDark} simulation suite using the \textsc{Rockstar} halo finder, i.e., using the same fitting algorithm. We compute the \cmr using the same pipeline and resolution limits as for our simulations. In particular, we use \textsc{Rockstar} halo catalogs for the {\it Bolshoi} (box size $250 \mpch$) and {\it MultiDark} ($1 \gpch$) simulations that model the same {\it WMAP7} cosmology as our reference simulations \citep{klypin_11, prada_12}. In addition, we use the {\it MDPL2}, {\it BigMDPL}, and {\it HMDPL} boxes that model the {\it MultiDark-Planck} cosmology with box sizes of $1$, $2.5$, and $4 \gpch$, respectively \citep{klypin_16}. These simulations are summarily referred to as the {\it MultiDark simulation suite}, and the catalogs are publicly available on the {\it Skies and Universes} website.

Figure~\ref{fig:comp_multidark} demonstrates that the differences largely remain when the same halo finder and fitting method are used, confirming that they are caused by the data from one or both of the simulation suites. In particular, the {\it Bolshoi/MultiDark} ({\it WMAP7}) results differ from ours in a mass-dependent fashion. The {\it MultiDark-Planck} \cmr agrees with our model to better than 10\% except at the highest halo masses at $z = 0$, where the simulation predicts no upturn. We have attempted to fit our model (Equation~\ref{equ:fit_func}) simultaneously to the \cmrs from our scale-free simulations and from the {\it MultiDark} suite. However, the best fit still deviates from the data by up to 15\%, largely because the high-$\nu$ predictions of {\it Bolshoi/MultiDark} and {\it MultiDark-Planck} are somewhat incongruent. We note that the \cmrs shown in Figure~\ref{fig:comp_multidark} are not the same as in \citet{klypin_11, klypin_16} and \citet{prada_12}. There, concentration was evaluated from the ratio of circular velocities at certain radii, leading to systematic differences \citep[e.g.,][]{meneghetti_13}. To avoid an unfair comparison, we refrain from showing the fitting functions from the original works in Figure~\ref{fig:comp_multidark}.

Our findings raise the question of how the simulated density profiles can differ so significantly. An obvious culprit could be the simulation code, given that {\it Bolshoi} and {\it MultiDark} were run with ART \citep{kravtsov_97, gottloeber_08} while our simulations and {\it MultiDark-Planck} were run with \textsc{Gadget2}. However, the different $N$-body algorithms have been compared and been found to agree relatively well \citep[e.g.,][]{knebe_00, diemand_04_convergence, heitmann_05, klypin_09}. 

More likely, the differences are caused by differences in the numerical parameters that determine the accuracy of $N$-body simulations, namely, mass resolution (the number of particles per halo), force resolution (the smoothing scale), and time stepping. As described in Section~\ref{sec:sims}, we have attempted to eliminate mass and force resolution effects by introducing a minimum number of particles per halo and by requiring the scale radius to be resolved by a sufficient number of force softening lengths \citep{moore_98, klypin_01}. We have carefully tested the convergence of our results with those parameters \citepalias{diemer_15}, and we note that the mass and force resolutions of some of our simulation boxes are rather similar to those of the {\it Bolshoi} simulation.

These arguments leave time stepping as the most likely culprit. $N$-body simulations can suffer from unphysical two-body scattering, which leads to artificial heating and thus lowers the central density of halos. The magnitude of this effect depends on the size of the time step \citep{knebe_00}. We cannot quantify the impact of time stepping without a detailed numerical study, but \citet{child_18} find that halving their time step shifts concentrations up by less than 5\% while doubling it lowers concentrations by about 8\% (their Appendix B2).

Comparing our time stepping to {\it Bolshoi/MultiDark} is difficult because ART and \textsc{Gadget2} use different algorithms to determine the time step. In particular, ART is an adaptive mesh refinement (AMR) code, meaning that the time step varies between resolution levels as $\Delta t \propto \rho^{-1/3}$. In \textsc{Gadget2}, it scales as $\Delta t = \sqrt{2 \eta \epsilon / |\vec{a}|}$, where $\eta$ is a free parameter and $\vec{a}$ is the acceleration vector \citep{springel_05_gadget2}. Around the scale radius where $\rho \propto r^{-2}$, this translates to $\Delta t \propto \rho^{-1/4}$ \citep{klypin_09}. However, it is not clear how these different time step scalings would translate into the mass trend shown in Figure~\ref{fig:comp_multidark}. To compare the time stepping in our simulations to {\it MultiDark-Planck}, we consider simulation boxes that have roughly the same mass resolution (L1000 and HMDPL, L0250 and MDPL). The {\it MultiDark-Planck} boxes use a force softening that is between $1.1$ and $1.6$ times smaller, an effect that is combined with a more aggressive time stepping parameter $\eta$ to result in about twice as many time steps as in our simulations. While this systematic difference suggests that {\it MultiDark-Planck} should be better converged than our simulations, there is no discernible overall offset in the concentrations, which is somewhat surprising given the findings of \citet{child_18}. Thus, if time stepping was the actual cause of the numerical differences, it would appear that the different time stepping algorithms in ART and \textsc{Gadget2} have a larger impact than the factor of two difference between {\it MultiDark-Planck} and our simulations.

We note that resolution and time stepping effects are not simply eliminated by arbitrarily decreasing the force smoothing scale and time step because the convergence depends on particle number, force resolution, and time stepping in a complicated fashion \citep{splinter_98, knebe_00, power_03, diemand_04_2body, joyce_09, joyce_13}. For example, even accurately integrated two-body collisions are unphysical and can alter the density profiles. Such effects could have escaped our convergence tests because mass, force, and time resolution are varied at the same time.

In summary, we find that our simulation data disagree with the {\it MultiDark} simulation suite by up to 15\%, likely due to the time stepping schemes used in the underlying suites of $N$-body simulations. This disagreement is of the same order as other systematic effects such as sample selection and the definition of concentration. We caution that the concentration of simulated halos is not a uniquely defined quantity and that it is likely affected by numerical inaccuracies. A detailed study of the impact of resolution on concentration will be undertaken in P. Mansfield et al. (2019, in preparation).


\section{The Importance of the Expansion Rate}
\label{sec:discussion}

One of the most important findings of this paper is that the physics of \cm evolution at low redshift cannot be encoded solely by the dependence on peak height and the effective slope of the power spectrum. Instead, it requires, at least, parameterization of an additional dependence on the cosmological expansion. We have argued that $\alphaeff$ is a physically motivated parameter that captures the late-time deviation from EdS  expansion in \LCDM, and that it appears to influence halo concentrations. In this section, we discuss future avenues for improving our understanding of this influence and of the capacity of $\alphaeff$ to describe it.

As discussed in Section~\ref{sec:theory:all_lcdm}, the dependence on cosmology could originate through both the evolving halo definition (which is unique in EdS up to the choice of the value of the constant $\Delta$) and the non-EdS linear growth factor. The former dependence is already explicitly included in our pseudo-evolution calculations and should be subdominant for physically accreting halos. Thus, our choice of $\alphaeff$ is motivated by the expected dependence on the evolution of $D(z)$.

There is also a separate physical motivation for the choice of $\alphaeff$ as in Equation~(\ref{equ:alphaeff}): a cosmology with a constant $\alphaeff = \alpha$ {\it different from unity} still corresponds to an EdS expansion law, $a \propto t^{2/3}$, and, taking power-law initial conditions, defines a broader (two-parameter) family of scale-free cosmologies than the one usually considered \citep{benhaiem_13, benhaiem_14}. In such cosmologies the universe expands faster than in ``standard'' ($\alpha = 1$) EdS for $\alpha < 1$ and slower for $\alpha > 1$, with a linear growth factor $D(z) = (1 + z)^{-\alpha}$. 

If the power spectrum is scale-free, we can generalize the derivation of the \cmr in the pseudo-evolving limit to include $\alpha$, and obtain
\begin{equation}
\frac{c}{\left[g(c)\right]^{\frac{1}{3-\gamma}}} = \frac{A(n,\alpha)}{ \nu^{1/\alpha}}  \,,
\label{cnu-pseudo-evolution-GSF}
\end{equation}
where $\gamma = 3(3+n)/(3+n+2\alpha)$. As shown in \citet{benhaiem_14}, the constant $\gamma$ is the exponent characterizing the decaying power-law behavior of the nonlinear two-point correlation function in the stable clustering approximation. This result is a  generalization of the corresponding one for the usual $\alpha = 1$ case, $\gamma = 3(3+n)/(5+n)$, which was originally derived by \citet{peebles_74}. The \cmr for the full mass range might take on a form similar to our fitting function of Equation~(\ref{equ:fit_func}), 
\begin{equation}
\frac{c}{ \left[g(c)\right]^{\frac{1}{3-\gamma}}}=  \frac{A(n,\alpha)}{\nu^{1/\alpha}} 
\left[ 1+\frac{\nu^{2}}{B(n,\alpha)} \right] \,.
\end{equation}
Such a parameterization would replace the $C(\alpha)$ factor and could be calibrated with large dedicated simulations of $\alpha$-universes, like those described  for modest sizes ($N=256^3$) in \citet{benhaiem_14}. An alternative avenue to further constrain the dependence on $\alphaeff$ would be to run \LCDM simulations far into the future where $\alphaeff \ll 1$.


\section{Conclusions}
\label{sec:conclusions}

We have presented an accurate, semianalytical model of the mean and median \cnur that describes simulation data over a wide range of cosmologies, masses, and redshifts. As intermediate products, we have developed a number of few-parameter fitting functions that are valid in certain limits, e.g., for low-mass halos or EdS cosmologies. Our model is publicly available through the \colossus code. Our main conclusions are as follows:
\begin{enumerate}
\item The assumption of pure pseudo-evolution (radius and mass changes compatible with a fixed density profile in physical coordinates) can quantitatively explain the behavior of the \cnur for $\nu \lsim 1.4$, in both \LCDM and scale-free cosmologies. 
\item While peak height and the effective power spectrum slope, $\neff$, can explain the majority of the evolution of the \cnur, we have shown that a third physical parameter is necessary: the effective exponent of linear growth, $\alphaeff$. The definition of $\neff$ matters when connecting scale-free and \LCDM cosmologies. Definitions based on $\sigma(R)$ fare better than those based on $P(k)$.
\item Including a dependence on $\alphaeff$, we have proposed a simple, semianalytical model with six free parameters (Equation~\ref{equ:fit_func}). This function fits our data to 5\% or better for virtually all tested cosmologies, halo masses, and redshifts. 
\item The predictions of our model are distinct from all previously proposed models. While they agree with the (updated) \citetalias{diemer_15} model to about 5\% for \LCDM cosmologies at $z < 6$, they diverge for scale-free cosmologies and at very high redshift.
\item Our model describes the \cmr in the {\it MultiDark} simulation suite to about 15\% accuracy. The disagreement is caused by numerical differences in the underlying $N$-body simulations, which are most likely due to the different time stepping schemes. While our simulations use fewer time steps than {\it MultiDark}, we cannot currently determine the relative convergence of the different simulation suites.
\end{enumerate}
While we have made progress in our quantitative understanding of the \cmr and its evolution at low halo masses, the physics that determines $c$ at high masses remain harder to quantify. In particular, the upturn due to unrelaxed halos is difficult to model and almost certainly depends on the exact technique of determining concentration. Similarly, we have introduced $\alphaeff$ as a physical variable, although our understanding of how $\alphaeff$ influences concentration is sorely lacking. Simulations of EdS universes with different values of $\alpha$ or far-future simulations of \LCDM cosmologies are needed to understand this dependence more systematically.


\vspace{0.5cm}

We are indebted to Philip Mansfield for extensive discussions regarding the impact of numerical effects on concentration. We thank Sownak Bose for comments on a draft of this paper. M.J. warmly thanks the Institute for Theory and Computation for hosting him as a sabbatical visitor for the academic year 2017-18, acknowledges David Benhaiem for collaboration and many discussions about halos in scale-free models, and thanks Ravi Sheth and Yan Rasera for numerous useful conversations on subjects related to this work. B.D. gratefully acknowledges the financial support of an Institute for Theory and Computation Fellowship. Support for program no. HST-HF2-51406.001-A was provided by NASA through a grant from the Space Telescope Science Institute, which is operated by the Association of Universities for Research in Astronomy, Incorporated, under NASA contract NAS5-26555.


\appendix

\section{Updated DK15 Parameters}
\label{sec:app:dk15update}

\begin{figure}
\centering
\includegraphics[trim =  0mm 19mm 2mm 85mm, clip, scale=0.76]{\figdir/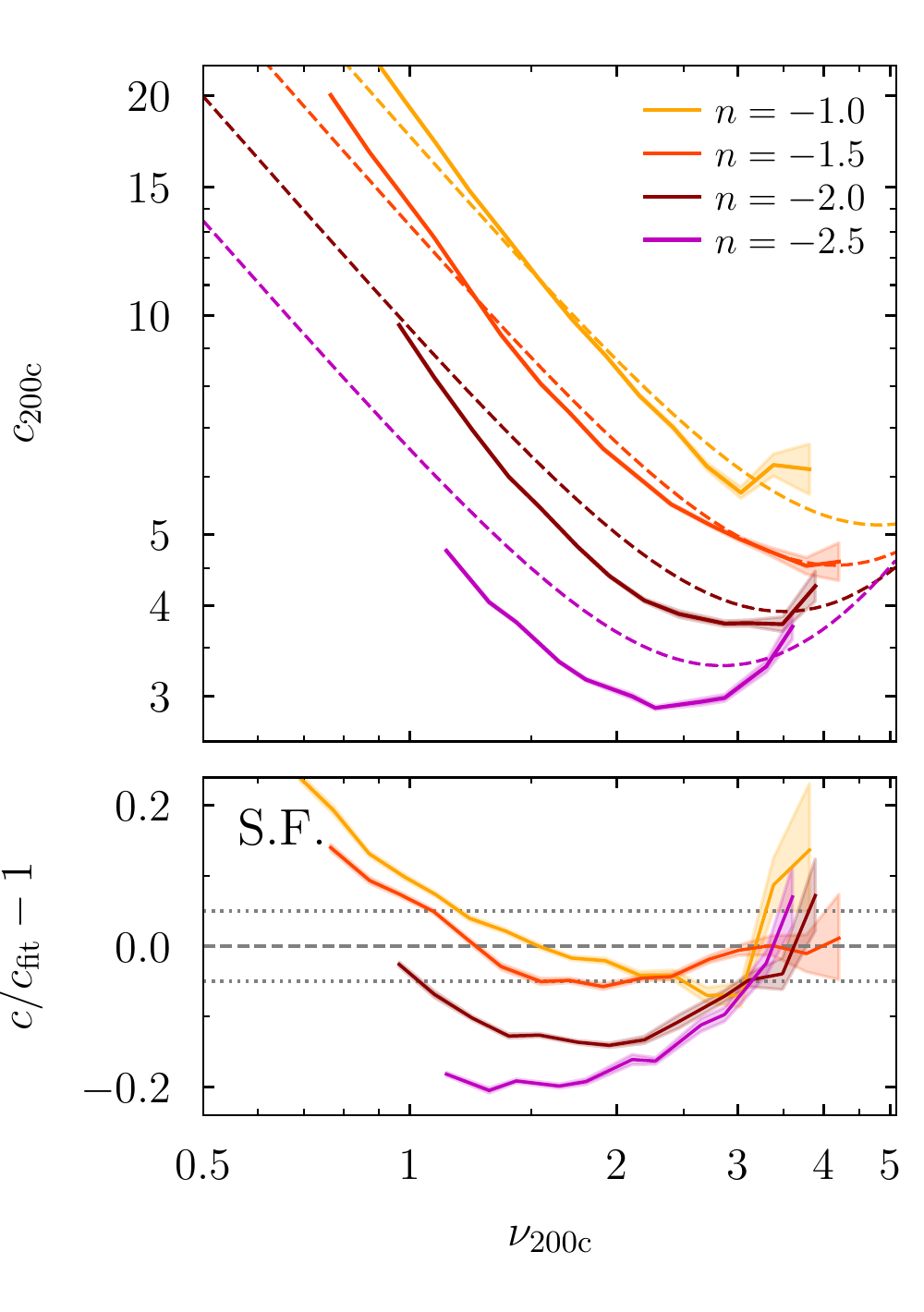}
\includegraphics[trim =  0mm 19mm 2mm 85mm, clip, scale=0.76]{\figdir/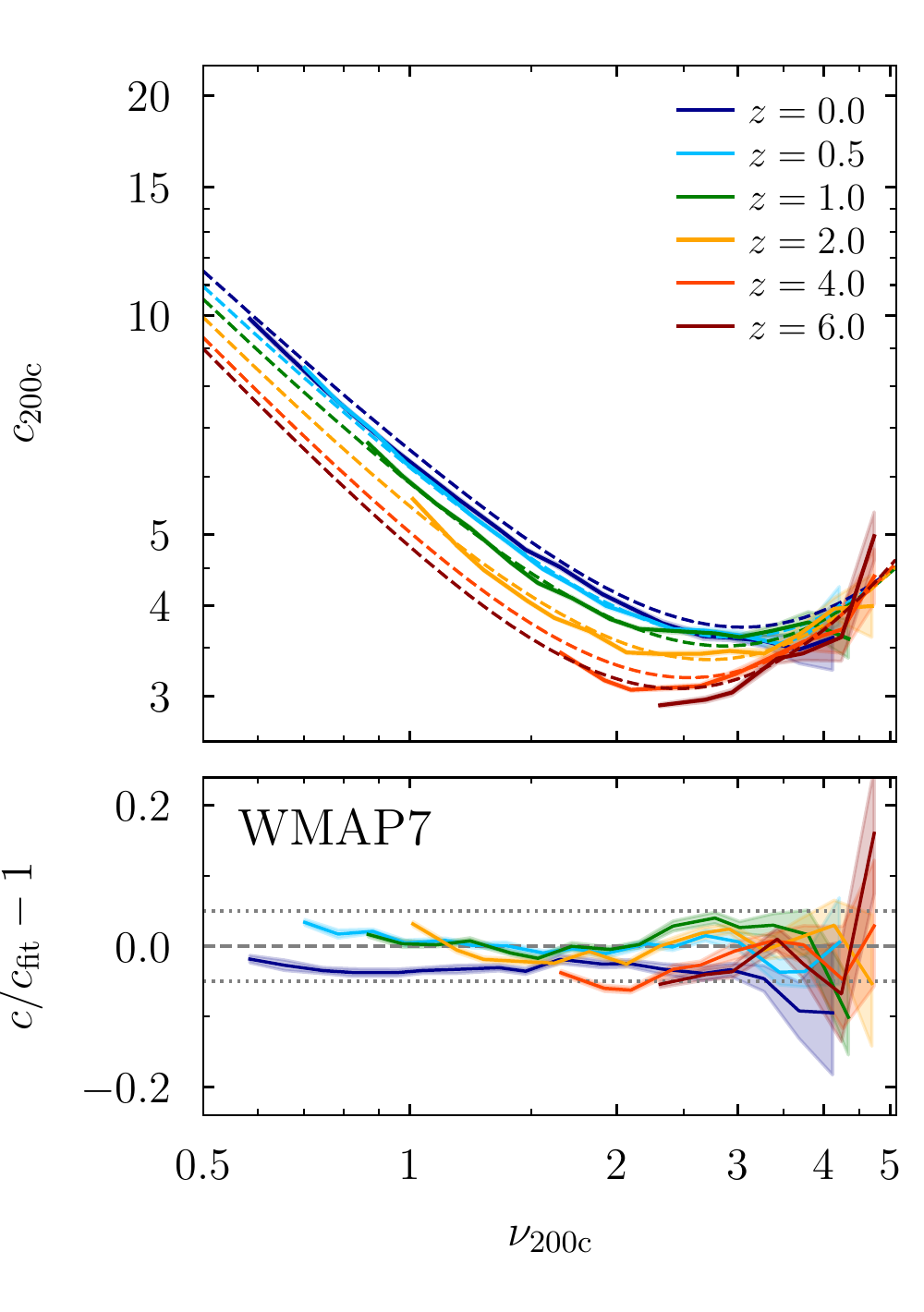}
\includegraphics[trim =  0mm  5mm 2mm 85mm, clip, scale=0.76]{\figdir/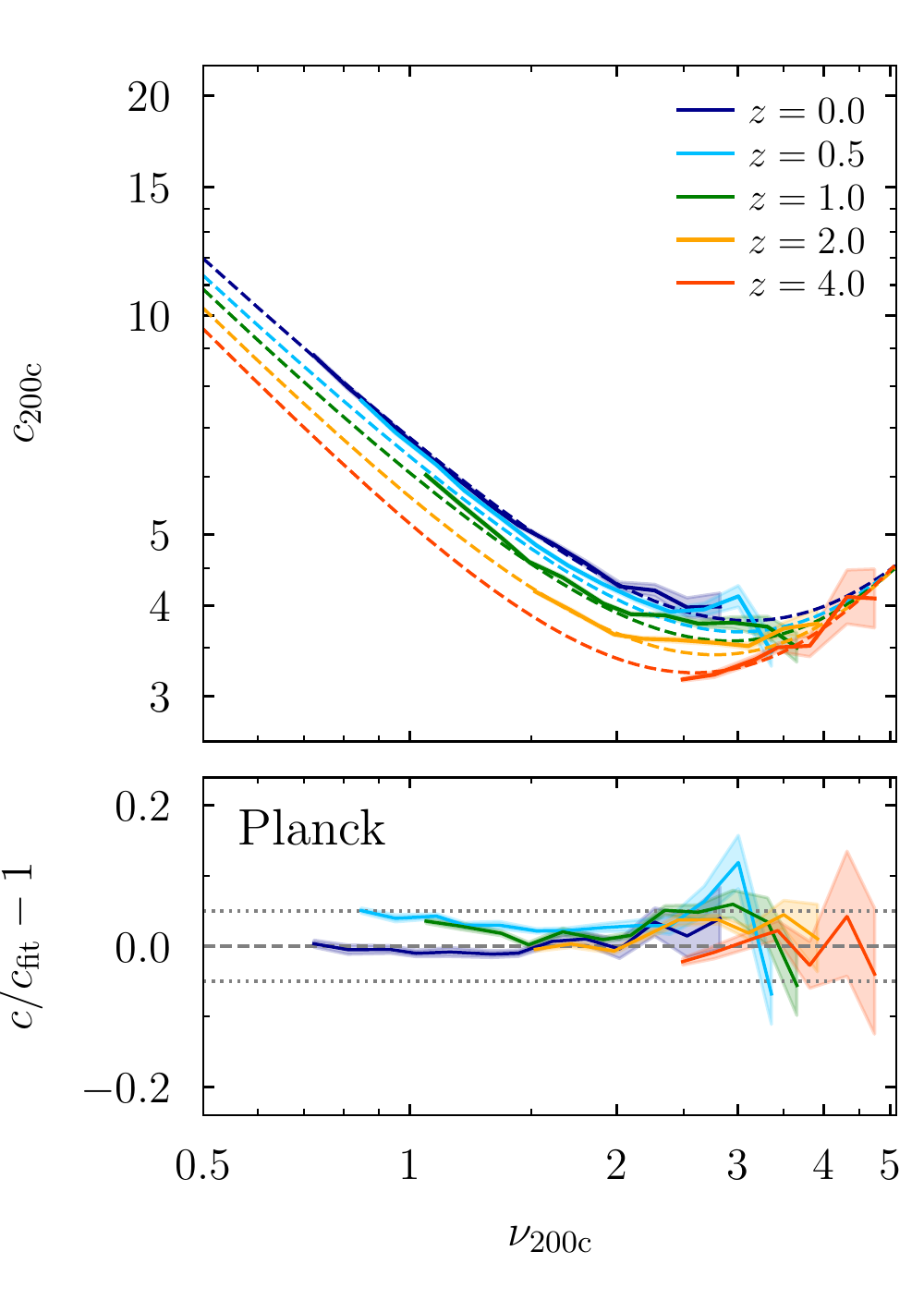}
\caption{Fit quality of the updated \citetalias{diemer_15} model. The panels show the relative difference between the simulation data (with a slight numerical error corrected compared to \citetalias{diemer_15}) and the recalibrated model (using the parameters in Table~\ref{table:dk15update}). The panels show the scale-free, {\it WMAP7}, and {\it Planck} cosmologies, from top to bottom. The colors and lines have the same meaning as in Figure~\ref{fig:main_fit}. The model still fits the \LCDM data to 5\% or better, but the fit to the scale-free models has become less accurate as a result of the changes in the \LCDM data. We discuss the reasons for this disagreement in Appendix~\ref{sec:app:dk15update}.}
\label{fig:dk15update}
\end{figure}

In the original \citetalias{diemer_15} data, a small numerical error meant that the $z > 0$ data of the {\it WMAP7} cosmology were miscalculated by a few percent, depending on redshift. We refit the \citetalias{diemer_15} model to the corrected data using the same weights on the different datasets (five times higher weight on \LCDM than on scale-free data, and twice that weight at $z = 0$). In contrast to \citetalias{diemer_15}, we introduce the same systematic error of 2\% as for the new model fits (Section~\ref{sec:sims}). The new best-fit parameters are listed in Table~\ref{table:dk15update} and have been implemented in the \colossus code (though the original parameter set is also available for backward compatibility). As implied by Figure~\ref{fig:model_comp_dk15}, the old and new models agree to 5\% or better at all redshifts and peak heights except at the highest peak heights, where the differences can reach 10\%. For the scale-free cosmologies, the differences are about 10\%. 

Figure~\ref{fig:dk15update} shows the fit quality of the new fit and can be directly compared to Figures 5--7 in \citetalias{diemer_15}. While the fit still matches the data to 5\% or better for the \LCDM cosmologies, a tension with the {\it WMAP7} data at $z = 0$ becomes apparent (which had been partially concealed by the numerical error in \citetalias{diemer_15}): the $z = 0$ concentrations are lower than expected from the model. In our new model, we have traced this tension to the effect of the non-EdS expansion rate, parameterized it using $\alphaeff$, and thus ensured a good fit at both low and high redshift. 

Moreover, we note that the fit to the scale-free models has become noticeably worse compared to \citetalias{diemer_15}. As in the original fit, the 
scale-free models were down-weighted by a factor of five compared to the \LCDM data, meaning that the worse match is a direct consequence of a 
tension between the \LCDM and scale-free simulations. In this work, we have resolved this tension through the definition of $\neff$: when using 
the slope of $\sigma(M)$ (Equation~\ref{neff-1}) rather than the slope of $P(k)$ (as in \citetalias{diemer_15}, Equation~\ref{neff-2}), the \LCDM fit 
remains almost the same while the scale-free fit improves by about 5\% in accuracy. Thus, we conclude that 
Equation~(\ref{neff-1}) is preferable to Equation~(\ref{neff-2}) in the sense that it provides a more physical match of $\neff$ to the unambiguous power spectrum slope in the scale-free models. For consistency, however, the parameters in Table~\ref{table:dk15update} refer to the slope of $P(k)$. For the \LCDM fits, the difference is negligible, and for the scale-free fits one should prefer the new model over that of \citetalias{diemer_15} because it provides a much more accurate fit.

\begin{deluxetable}{lccl}
\tablecaption{Updated DK15 parameters}
\tablehead{
\colhead{Par.} &
\colhead{Median } &
\colhead{Mean } &
\colhead{Explanation}
}
\startdata
\rule{0pt}{1ex} $\kappa$   & $1.00$   & $1.00$  & Loc. in $k$ where $n$ is computed \\
\rule{0pt}{0pt} $\phi_0$   & $6.58$   & $6.66$  & Normalization of $c$ floor \\
\rule{0pt}{0pt} $\phi_1$   & $1.27$   & $1.37$  & Slope dependence of $c$ floor \\
\rule{0pt}{0pt} $\eta_0$   & $7.28$   & $5.41$  & Norm. of $\nu$ where $c$ is minimum \\
\rule{0pt}{0pt} $\eta_1$   & $1.56$   & $1.06$  & Slope dep. of $\nu$ where $c$ is min. \\ 
\rule{0pt}{0pt} $-\alpha$  & $-1.08$  & $-1.22$ & Slope of \cnur at low $\nu$ \\
\rule{0pt}{0pt} $\beta$    & $1.77$   & $1.22$  & Slope of \cnur at high $\nu$
\enddata{}
\tablecomments{Updated best-fit parameters for the \citetalias{diemer_15} model. These parameters replace those given in Table~3 of \citetalias{diemer_15} which were slightly off owing to a numerical error.}
\label{table:dk15update}
\end{deluxetable}

\section{Conversion to Other Mass Definitions}
\label{sec:app:mdef}

When fitting Equation~(\ref{equ:fit_func}) to mass definitions other than $\ctoc$, we find that the fit degrades. For example, when fitting $\ctom$, the differences increase to 10\% for \LCDM and 20\% for the scale-free cosmologies. While the original expression our model was based on, Equation~(\ref{equ:c_pe_general}), includes the effects of mass definition, we have given up that generality when we introduced a phenomenologically motivated parameterization at the high-mass end.

Instead of attempting to find expressions for each mass definition, we follow \citetalias{diemer_15} in computing $\ctoc$ and converting it to other definitions assuming a fixed NFW profile (their Appendix C). Like \citetalias{diemer_15}, we find that the conversion to $\cvir$ and $\ctom$ maintains a fit accuracy of 5\% up to $\nu \lsim 2.5$. At higher $\nu$, the conversion overestimates the low-redshift concentrations by up to 15\%. For higher overdensities such as $\cfoc$, the conversion overestimates the true concentration above $\nu = 2$, increasing to a 15\% difference at $\nu = 4$. The conversions are automatically performed by the \colossus code.

\bibliographystyle{aasjournal}
\bibliography{/Users/benedito/University/Docs/_LatexInclude/sf}

\end{document}